\documentclass[epj]{svjour}

\usepackage{epsfig,multicol,bbm,physics,lscape,multirow}
\usepackage{amsmath}
\usepackage{amssymb}
\usepackage{subfigure}
\usepackage{threeparttable}
\usepackage{graphicx}
\usepackage{url}

\def\boostauthor[#1]#2{{#2}$^{\, #1}$}
\def\specialboostauthortwo[#1][#2]#3{{#3}$^{\, #1, #2}$}
\def\specialboostauthorthree[#1][#2][#3]#4{{#4}$^{\, #1, #2, #3}$}
\def\boosteditor[#1]#2{{#2}$^{\, #1, *}$}
\def\boostaffiliation[#1]#2{$^{#1\,}$ {#2}\\}

\begin{document}

\title{Boosted objects: a probe of beyond the standard model physics~\footnote{Report prepared by the hadronic working group of the BOOST2010 workshop at the University of Oxford} }

\author{ A. Abdesselam \inst{1} \and A. Belyaev \inst{2,3} \and
  E. Bergeaas Kuutmann \inst{4} \and U. Bitenc \inst{5} \and
  G. Brooijmans \inst{6} \and J. Butterworth \inst{7} \and P. Bruckman
  de Renstrom \inst{8} \and D. Buarque Franzosi \inst{9} \and
  R. Buckingham \inst{1} \and B. Chapleau \inst{10} \and M. Dasgupta
  \inst{11} \and A. Davison \inst{7} \and J. Dolen \inst{12} \and
  S. Ellis \inst{13} \and F. Fassi \inst{14} \and J. Ferrando \inst{1}
  \and M.T. Frandsen \inst{15} \and J. Frost \inst{16} \and T. Gadfort
  \inst{17} \and N. Glover \inst{18} \and A. Haas \inst{19} \and
  E. Halkiadakis \inst{20} \and K. Hamilton \inst{21} \and C. Hays
  \inst{1} \and C. Hill \inst{22} \and J. Jackson \inst{3} \and
  C. Issever \inst{1} \and M. Karagoz \inst{1} \and A. Katz \inst{23}
  \and L. Kreczko \inst{24} \and D. Krohn \inst{25} \and A. Lewis
  \inst{1} \and S. Livermore \inst{1} \and P. Loch \inst{26} \and
  P. Maksimovic \inst{27} \and J. March-Russell \inst{15} \and
  A. Martin \inst{28} \and N. McCubbin \inst{3} \and D. Newbold
  \inst{24} \and J. Ott \inst{29} \and G. Perez \inst{30} \and
  A. Policchio \inst{13} \and S. Rappoccio \inst{27} \and A.R. Raklev
  \inst{31} \and P. Richardson \inst{18} \and G.P. Salam
  \inst{25,32,33} \and F. Sannino \inst{34} \and J. Santiago \inst{35}
  \and A. Schwartzman \inst{19} \and C. Shepherd-Themistocleous
  \inst{3} \and P. Sinervo \inst{36} \and J. Sjoelin \inst{37,38} \and
  M. Son \inst{39} \and M. Spannowsky \inst{40} \and E. Strauss
  \inst{19} \and M. Takeuchi \inst{41} \and J. Tseng \inst{1} \and
  B. Tweedie \inst{27,42} \and C. Vermilion \inst{43} \and J. Voigt
  \inst{29} \and M. Vos \inst{44} \and J. Wacker \inst{19} \and
  J. Wagner-Kuhr \inst{29} \and M.G. Wilson \inst{19}}

\institute{University of Oxford, Department of Physics, Denys Wilkinson Building, Keble Road, Oxford, OX1 3RH, UK 
\and School of Physics \& Astronomy, University of Southampton, Highfield, Southampton SO17 1BJ, UK
\and Rutherford Appleton Laboratory, Science and Technology Facilities Council, Harwell 
Campus, Didcot OX11 0QX, UK
\and Deutsches Elektronen-Synchrotron, DESY, location Zeuthen, Platanenallee 6, D-15738 Zeuthen, Germany 
\and Albert-Ludwigs-Universit\"at Fak. f\"ur Mathematik und Physik, Hermann-Herder Str. 3, D-79104 Freiburg i.Br., Germany 
\and Columbia University, Nevis Laboratory, 136 So. Broadway, Irvington, NY 10533, USA 
\and Department of Physics and Astronomy, University College London, WC1E 6BT, UK 
\and Institute of Nuclear Physics P.A.N., ul. Radzikowskiego 152, 31-342 Krakow, Poland
\and Universit\`{a} degli Studi di Torino, Dipartimento di Fisica Teorica, Via Pietro Giuria 1, 10125 Turin, Italy
\and McGill University, High Energy Physics Group, 3600 University Street, Montr\'{e}al, Qu\'{e}bec H3A 2T8, Canada 
\and School of Physics and Astronomy, University of Manchester, Manchester, M13 9PL, UK 
\and University of California, Davis, Davis, CA 95616, USA 
\and Department of Physics, University of Washington, Box 351560, Seattle, WA 98195, USA 
\and CNRS/CC-IN2P3, 43 Bd du 11 Novembre 1918 69622 Villeurbanne, France 
\and Dalitz Institute for Theoretical Physics, Department of Physics, University of Oxford, Oxford, OX1 3RH, UK 
\and Cavendish Laboratory, University of Cambridge, J.J. Thomson Avenue, Cambridge, CB3 0HE, U.K. 
\and Brookhaven National Laboratory, Physics Department, Upton, NY 11973,USA 
\and Institute of Particle Physics Phenomenology, Department of Physics, University of Durham, Durham, DH1 3LE, UK
\and SLAC National Accelerator Laboratory, Menlo Park, CA 94025, USA
\and Rutgers University, Department of Physics and Astronomy, 136 Frelinghuysen Road, Piscataway, NJ 08854, USA
\and INFN, Sezione di Milano-Bicocca, Piazza della Scienza 3, 20126 Milan, Italy
\and Department of Physics, the Ohio State University, 191 W. Woodruff Ave., Columbus, OH 43210, USA  
\and Department of Physics, University of Maryland, College Park, MD 20742, USA
\and H. H. Wills Physics Laboratory, University of Bristol, Bristol BS8 1TL, UK
\and Department of Physics, Princeton University, Princeton, NJ 08544, USA     
\and Department of Physics, University of Arizona, Tucson, AZ 85719, USA
\and Department of Physics and Astronomy, Johns Hopkins University, 3400 N. Charles St., Baltimore, MD 21218, USA
\and Fermi National Accelerator Laboratory, Batavia, IL, 60510, USA    
\and KIT, Institut f\"ur Experimentelle Kernphysik, 
Wolfgang-Gaede-Str. 1 76131, Karlsruhe, Germany
\and Department of Particle Physics and Astrophysics, Weizmann Institute, 76100 Rehovot, Israel     
\and Department of Physics, University of Oslo, P.O. Box 1048 Blindern, N-0316 Oslo, Norway
\and LPTHE, UPMC Univ.~Paris 6 and CNRS UMR 7589, Paris, France
\and Department of Physics, Theory Unit, CERN, CH-1211 Geneva 23, Switzerland 
\and Center for Particle Physics Phenomenology, CP3-Origins, University of Southern Denmark, Odense M. Denmark  
\and CAFPE and Depto. de Fisica Teorica y del Cosmos, U. of Granada, E-18071 Granada, Spain    
\and Department of Physics, University of Toronto, 60 Saint George Street, Toronto, M5S 1A7, Ontario, Canada          	
\and Oskar Klein Centre, Department of Physics, Stockholm University, SE-10691 Stockholm, Sweden
\and Department of Physics, Stockholm University, SE-10691 Stockholm, Sweden
\and Department of Physics, Yale University, New Haven, CT 06511, USA    
\and Institute of Theoretical Science, University of Oregon, Eugene, OR 97403-5203, USA
\and Institute for Theoretical Physics, Uni Heidelberg, Philosophenweg 16,  D-69120 Heidelberg, Germany
\and Physics Department, Boston University, Boston, MA 02215, USA
\and Department of Physics \& Astronomy, University of Louisville, Louisville, KY 40292, USA
\and Instituto de F\'isica Corpuscular, IFIC/CSIC-UVEG, PO Box 22085, 46071 Valencia, Spain    
}

\abstract{We present the report of the hadronic working group of the BOOST2010 
workshop held at the University of Oxford in June 2010. The first part contains
a review of the potential of hadronic decays of highly boosted particles as an 
aid for discovery at the LHC and a discussion of the status of tools developed 
to meet the challenge of reconstructing and isolating these topologies. In the 
second part, we present new results comparing the performance of jet grooming 
techniques and top tagging algorithms on a common set of benchmark channels. 
We also study the sensitivity of jet substructure observables to the 
uncertainties in Monte Carlo predictions.}

\authorrunning{M. Karagoz, G. P. Salam, M. Spannowsky, M. Vos (editors)}
\titlerunning{Boosted objects: a probe of BSM physics}

\maketitle








\section{Introduction}
\label{introduction}

The LHC has started to explore the multi-\tev regime. The production of presently unknown particles is perhaps the most exciting prospect for the general purpose experiments ATLAS~\cite{atlas} and CMS~\cite{cms}. In both experiments, searches for Physics Beyond the Standard Model form a key element of the rich physics programme.

At the LHC, many of the particles that we considered to be heavy at previous accelerators will be 
frequently produced with a (transverse) momentum greatly exceeding their rest mass. 
Good examples are the electro-weak gauge bosons $W^{\pm}$ and $Z^{0}$, the top quark, 
the Higgs boson or bosons and possibly other new particles in the same mass range. 
The abundant presence of heavy SM particles will yield promising signatures for searches for physics Beyond the Standard Model (BSM physics).
When these boosted objects decay they form a highly collimated topology in the detector. 
Algorithms and techniques developed for the reconstruction and isolation of objects produced at rest are often inadequate for their boosted counterparts. New tools must be developed to fully benefit from the potential of these states.

In recent years, a fruitful dialogue has developed between theorists and the LHC and Tevatron experimentalists. A number of workshops have fuelled collaboration in the investigation of new signatures and the development of experimental techniques. The series started with the BOOST09~\cite{boost09} workshop at Stanford National Accelerator Centre (SLAC) and continued at the Jet Substructure workshop at the University of Washington~\cite{seattle} in January 2010. From the 22nd to the 25th of June of 2010, BOOST2010~\cite{boost2010} was held at the University of Oxford.

At the BOOST2010 workshop, two working groups were set up to concentrate on the leptonic and hadronic decays of boosted objects. Mixed decays to quarks and leptons (i.e. top decay to $ W^{\pm} b $ followed by $ W^{\pm} \rightarrow l^{\pm} \nu_l$) were also covered by the hadronic working group. Both working groups met in several parallel sessions during the workshop and organised follow-up meetings in the subsequent months. In this paper we present the report of the hadronic working group.

Hadronic boosted objects have received considerable attention recently and the available literature is steadily increasing. We start this report with three brief sections that provide a review of the most important developments. 

Many groups have studied the phenomenology of boosted hadronic topologies, discovering novel ways of performing SM measurements and BSM searches. In section~\ref{sec:had:models} we present a review of the results published to date.

The reconstruction of hadronic decays of boosted $W, Z$ bosons and top quarks (and new particles with similar mass) is particularly challenging. The partons 
formed in the decay are typically too close to be resolved by a jet 
algorithm\footnote{To be quantitative, consider the following rule of 
thumb for a two-body decay: To resolve the two partons of a $ X \rightarrow q {\bar{q}}$ decay, a radius (or more generally a jet size) of $R < 2 m_X / p_T$ must be chosen. For $ p_T \gg m_X $, the value for $R$ must be chosen exceedingly small. For $m_X= $ 80~\gev, the minimal $R$ is equal to 0.4 for a transverse momentum of 400~\gev. To set the scale: 95 \% of the energy in a 400~\gev jet is contained in a jet of size 0.4~\cite{atlasjetshapes,cmsjetshapes}.}. 
In this case only an analysis of the substructure of the {\em fat-jet} can reveal its heavy-particle origin. We give an overview, in section~\ref{sec:had:tools}, of the increasingly sophisticated tools developed for this purpose. 

In the final review section, Section ~\ref{sec:had:exp}, we present a brief review of the experimental status of jet substructure in past experiments. 

In the sections thereafter, we present new results obtained in studies initiated during the workshop. In section~\ref{sec:had:samples}, we present the Monte Carlo samples generated in the hadronic working group. We make these samples available to serve as a benchmark test for the performance of new techniques. 

In section~\ref{sec:had:comp_groom}, we return to the jet grooming techniques 
introduced in section~\ref{sec:had:tools}. We present an estimate of their 
performance on the benchmark samples. 

Jet substructure may be subject to considerable uncertainties in the predictions of popular Monte Carlo models. The sensitivities of the most important observables to variations in the parton shower model, the underlying event and detector effects are investigated in section~\ref{sec:had:robust}.

Finally, we compare the performance of several top-tagging algorithms in section~\ref{sec:had:comp}.

 The review sections, benchmark samples and results are intended to foster the exciting new developments that boost our discovery potential. We hope that this report be an incentive for further work and in particular for studies of the substructure of highly energetic jets in the earliest LHC data.














\section{Models and signatures}
\label{sec:had:models}
  



Many groups have studied the phenomenology of boosted hadronic decays, considering two major sources: 
\begin{itemize}
\item In BSM physics scenarios, where a heavy resonance 
decays to particles with intermediate masses that then decay to light
quarks, i.e. ${X}_{heavy} \rightarrow {Y}_{interm.}
\rightarrow \rm{jets}$, where $X_{heavy}$ is an unknown heavy
resonance and \\ $Y_{interm.}$ may be a SM particle with intermediate
mass ($W$, $Z$, top) or a BSM particle. In scenarios where $m_X \gg m_Y$ the intermediate particles are naturally boosted (i.e. $\pt \gg m$).
\item 
Even if only a relatively small fraction of signal events are produced with large
transverse momentum, focusing on those events can be a superior way of
disentangling the signal from the backgrounds.
Phenomenological studies~\cite{Butterworth:2008iy,ATLASHV,Butterworth:2009qa,Plehn:2009rk,Soper:2010xk,Brooijmans:2010tn} have indeed shown that, because of 
the kinematic features, event reconstruction efficiencies, $b$-tagging 
efficiencies~\cite{ATLASHV,giacinto} and the jet energy resolution can be 
improved and that combinatorial problems in the identification of the decay 
products of ${Y}_{interm.}$ are reduced.
\end{itemize}

Without going into the details of each study, in this section we 
briefly review which specific scenarios have been addressed in phenomenological 
studies. We establish the extent to which the use of the techniques of
section~\ref{sec:had:tools} have been shown to increase 
the discovery potential of the LHC experiments. The four subsections
correspond to four categories of boosted objects: boosted Higgs 
bosons, boosted top quarks, electroweak gauge bosons and finally boosted 
BSM particles.

\subsection{Boosted Higgs bosons}

The main purpose of running the LHC is to understand the nature of electroweak
symmetry breaking, e.g. by confirming or modifying the minimal
mechanism of the SM. The detection of a light SM Higgs
boson ($m_H<130$~\GeV) is particularly difficult and until recently relied
mainly on two channels: The dominant gluon-initiated production
mechanism followed by Higgs
decay to photons, $pp \rightarrow H \rightarrow \gamma \gamma$, or 
production through vector boson fusion followed by Higgs decay to 
$\tau$ leptons, $pp \rightarrow Hjj \rightarrow jj\tau^- \tau^+$. 

Butterworth, Davison, Rubin and Salam \cite{Butterworth:2008iy} (BDRS)
studied the case of a 
light Higgs boson ($m_H\sim 120$~\GeV) produced
in association with an electroweak gauge boson. The leptonic decay of 
the associated vector boson provides an efficient trigger for these events. 
Cuts on the leptonic decay products ensure that the 
electroweak gauge boson and the recoiling Higgs boson are produced with 
a large transverse momentum. The Higgs boson decays into a collimated 
$b \bar{b}$ pair with a large branching fraction. The analysis employs the 
Cambridge/Aachen (C/A) jet algorithm~\cite{Dokshitzer:1997in,Wobisch:1998wt} 
to investigate the jet substructure of the single, merged jet produced by the 
two $b$ quarks. 
This study demonstrated that $VH$ production can be a discovery
channel, with a significance $S/\sqrt{B} \simeq 4.5$, 
assuming $30~\mathrm{fb}^{-1}$ of data and $\sqrt{s}=14$~\TeV.
The ATLAS collaboration was able to reproduce this study in a 
full-detector simulation with only slightly smaller statistical 
significance~\cite{ATLASHV}.  


The $ZH$ channel was used by Soper and Spannowsky \cite{Soper:2010xk} 
to show that the combined use of the jet grooming techniques discussed in 
Section~\ref{sec:had:tools} can improve the 
confidence level of a Higgs detection.

One of the major discovery channels for a light SM Higgs boson in
ATLAS and CMS reports~\cite{:1999fr,CMStth2,CMStth} was 
the $t\bar{t}H$ production channel with 
subsequent Higgs decay to bottom quarks. Further studies revealed a
very poor signal-to-background ratio of $1/9$~\cite{camminschumacher,Benedetti:2007sn}, making the channel very
sensitive to systematic uncertainties which might prevent it
from reaching a $5\sigma$ significance for any luminosity. However, at high transverse momentum,
after reconstructing
the boosted, hadronically decaying top quark as well as the Higgs boson and
requiring 3 $b$-tags, Plehn, Salam and Spannowsky \cite{Plehn:2009rk} find a 
signal-to-background ratio of roughly 1/2, while keeping $S/\sqrt{B}$ at 
a similar value to that in Ref.~\cite{camminschumacher}.

For some scenarios with non-SM decays, the Higgs boson may evade the
constraints from the LEP experiments. 
Two recent studies of a light Higgs boson \\ ($m_H\sim 80-120$~\GeV)
in the vector boson associated ($VH$) and $t\bar{t}H$ channels have used
subjet techniques to ``un-bury'' its decay, via two pseudo-scalars, 
to four gluons from the large QCD backgrounds \cite{Chen:2010wk,Falkowski:2010hi}.

The reconstruction of a boosted Higgs boson from $H\rightarrow b\bar{b}$
in BSM models has been discussed in Ref.~\cite{Kribs:2009yh} 
and~\cite{Butterworth:2007ke}, with a specific
application to cascade decays in the Minimal 
Supersymmetric Standard Model (MSSM)~\cite{Kribs:2010hp}. For the
lightest CP-even Higgs boson, a high
statistical significance has been found with as little as $10~\rm{fb}^{-1}$ 
at \\ \mbox{$\sqrt{s}=14$~\TeV}, provided 
its production from neutralino or chargino decays is not too rare.


\subsection{Boosted top quarks}

The reconstruction of boosted top quarks was one of the first 
applications of subjet techniques. Due to the three-pronged decay of a 
hadronically decaying top quark and the known $W$ boson mass, the 
radiation pattern of its decay products provides strong discriminating 
power from QCD-induced light parton jets. Many different top-tagging 
approaches have been proposed, mainly focusing on scenarios where the 
boosted tops are decay products of much heavier resonances. These top quarks 
are naturally boosted ($p_{T_{\rm{top}}}>500$~\GeV).

The tools and algorithms proposed for top-tagging are discussed in 
more detail in Section~\ref{sec:had:tools} and the performance of
these algorithms is discussed in Section~\ref{sec:had:comp}; here we 
only review briefly some uses in the literature.

Early ATLAS studies used the so-called ``YSplitter'' 
tool~\cite{brooijmans}, proposed in \cite{Butterworth:2002tt} and further detailed in \cite{butterworth}, for the 
identification of hadronically decaying top quarks. These studies 
established that hadronic decays of boosted top quarks can be identified 
efficiently amongst the background from QCD jet production. In a later 
study~\cite{brooijmans2}, a likelihood analysis was used to improve the tagging 
efficiency further.  
 


The so-called ``Hopkins'' top-tagger of Kaplan, Rehermann, Schwartz
and Tweedie~\cite{Kaplan:2008ie} has been used in phenomenological studies~\cite{Bhattacherjee:2010za} and full detector simulation studies 
have been performed by CMS with a slightly modified 
implementation~\cite{cmsnote3}.
In a further CMS study~\cite{cmsnote2} this tagger was applied 
to the $t\bar{t}$ all-hadronic channel, showing a very high background 
rejection while keeping a reasonable signal efficiency for high 
\pt{} objects. Using this tagger the C/A algorithm 
outperforms both the $k_T$ or anti-$k_T$ algorithms in reconstructing
the top~\cite{cmsnote3}.


Taggers to reconstruct hadronic tops with moderate transverse 
momentum ($p_T \gtrsim 200$~\GeV), 
proposed by Plehn, Salam, Spannowsky and Takeuchi~\cite{Plehn:2009rk,Plehn:2010st}, were used in Ref.~\cite{Plehn:2010st} to reconstruct the light 
top squark of the MSSM in a final-state with only jets and missing 
transverse energy.



Top quarks are predominantly produced in pairs at hadron colliders. 
The lepton + jets and fully leptonic final states are easier to isolate from the QCD 
background at hadron colliders than the fully hadronic final state. 
Non-boosted top quarks decaying to $b \ell \nu_l$ provide 
an isolated charged lepton suitable for triggering, large missing 
transverse energy and a $b$ quark: three good handles for reducing 
the QCD background.

For boosted top quarks decaying leptonically, however, 
QCD jet production may again be a dangerous 
background. The rejection achieved by flavour tagging is severely degraded 
for very high $p_T$ jets~\cite{vos} and the $E_T^{miss}$ resolution may be
insufficient to reveal the presence of the relatively low $p_T$ neutrino. 
Finally, the lepton from $W^{\pm}$ boson decay and the $b$-jet 
often merge and traditional lepton isolation criteria result in a
significant loss of signal efficiency. 

In Ref.~\cite{Thaler:2008ju},
several alternatives to lepton isolation were
proposed with better performance in lepton + jets events.
Rehermann and Tweedie \cite{Rehermann:2010vq} propose a ``mini-isolation'' cut at 
the tracker level for the lepton, which results 
in a very high background rejection rate.


A full simulation study~\cite{ATL-PHYS-PUB-2010-008}
investigated the sensitivity of the ATLAS experiment to resonant production
of $ t \bar{t} $ pairs. The lepton + jets final state was selected 
using a combination of the leptonic observables developed in 
Ref.~\cite{Thaler:2008ju} and~\cite{Rehermann:2010vq} and a 
hadronic top-tagger based on an evolution of the work of 
Ref.~\cite{brooijmans2} and Ref.~\cite{Thaler:2008ju}. 
This study was specifically aimed at early ATLAS data ($200$~pb$^{-1}$ at 
$\sqrt{s}=10$~\TeV{} or $1$~fb$^{-1}$ at $\sqrt{s}=7$~\TeV{}) and the 
algorithm was
adapted to perform well for tops with only moderate boost. Its performance 
was found to compare favourably with that of a more traditional approach 
for a resonance mass as low as 1~\tev, showing that boosted tops 
are an interesting probe for new physics even in the earliest stages of
the experiment.


The CMS collaboration also investigated top pair production in the muon+jets decay channel 
($t\bar{t}\rightarrow \mu\nu b~bqq'$)~\cite{cmsnote,cmsnote4}. 
In both CMS analyses, 
jets were reconstructed with the SISCone algorithm~\cite{Salam:2007xv} and 
no top-tagging algorithm had been applied for the hadronically decaying top 
quark. Instead, either only the lepton isolation criterion had been 
relaxed~\cite{cmsnote4} or both the lepton isolation and 
the number of reconstructed jets criteria had been 
relaxed~\cite{cmsnote}. To estimate background from QCD multi-jet events, a 
data driven method was developed. 
While the method described in~\cite{cmsnote4} is more focused on a good mass resolution, it is suitable for searches for massive $t\bar{t}$ resonances in the lower end of the mass spectrum (around 1 \TeV) in the very early stages of data-taking. The analysis of Ref.~\cite{cmsnote} takes the boosted topology of the decay products into account and achieves significantly better cross section limits for massive resonances (2-3~\TeV).



\subsection{Boosted electro-weak gauge bosons}

Longitudinal vector boson scattering can help to unravel the nature of 
electroweak symmetry breaking, in particular, if no Higgs is found.
In the vector boson fusion process, the two vector bosons are produced in association with 
two forward jets and tend to be central and have high transverse momentum  
in the kinematic region relevant to boosted studies. 
Allowing one of the two vector bosons 
to decay to quarks enhances the branching ratio and subjet techniques 
can be used to reconstruct this 
vector boson from the collimated decay products. 
This was one of the first applications of jet substructure by
Butterworth, Cox and Forshaw \cite{Butterworth:2002tt}, 
and was further treated in~\cite{Han:2009em}, including the use of 
polarization. 

Building on Ref.~\cite{Butterworth:2002tt}, the ATLAS simulation study described in 
Ref.~\cite{CSC-exotics-VBS}, investigated a chiral Lagrangian model, 
in which a scalar or a vector resonance decays to two vector bosons that 
decay semi-leptonically. The hadronically decaying vector 
boson is found by investigating the jet mass of $k_T$ jets~\cite{kt,kt2} 
with a jet-size parameter of \mbox{$R=0.6$}. By also considering the specific 
phenomenology of vector boson scattering, i.e. the presence of 
high-$|\eta|$ jets and the absence of other central jets due to the lack of colour 
flow between the initial protons, it is shown that a semi-leptonically 
decaying $800$~\GeV{} $WZ$ resonance with a production cross section of 
$0.65$~fb can be discovered in 60~fb$^{-1}$ of integrated luminosity
at $\sqrt{s}=14$~\TeV. 

The reconstruction of boosted electroweak gauge bosons can also be used 
in SUSY cascade decay chains, as shown by Butterworth, Ellis and 
Raklev~\cite{Butterworth:2007ke}, 
to obtain information about the masses and branching ratios of SUSY 
particles. 

Searches for a heavy Standard Model Higgs boson focus on the so-called
`gold plated mode' where the Higgs decays to two leptonic $Z$
bosons which are naturally boosted. 
By requiring one of the $Z$ bosons to decay hadronically, Hackstein
and Spannowsky~\cite{Hackstein:2010wk} found the semi-hadronic channel to be 
at comparable significance  to the purely leptonic channel for detecting 
the Higgs boson. 
A combination of several subjet techniques was deployed, as suggested
in Ref.~\cite{Soper:2010xk}, to reconstruct the boosted, hadronically 
decaying $Z$ boson. 
Assuming the existence of a chiral fourth generation, the Higgs boson could 
be detected or ruled out with only $1~\rm{fb^{-1}}$  of data at 
$\sqrt{s}=7$~\TeV{} in this channel. Englert, Hackstein and 
Spannowsky~\cite{Englert:2010ud} showed that the semi-leptonic channel 
also provides sensitivity to the CP property of a heavy scalar resonance.

Only recently Katz, Son and Tweedie~\cite{Katz:2010mr} studied boosted 
electroweak bosons, e.g. $W$, $Z$ or Higgs, from a $Z'$ resonance. They showed
that reconstructing these bosons using the BDRS approach yields
significant improvements for the $Z'$ discovery potential compared to
previous analysis.

\subsection{Boosted BSM physics particles}

Currently, the published literature 
on the reconstruction of boosted BSM particles with unknown mass is fairly
sparse. One 
exception is the study from Butterworth, Ellis, Raklev and 
Salam~\cite{Butterworth:2009qa}. 
If baryon-number-violating couplings are present in supersymmetry, the 
lightest neutralino can decay into three quarks. These 
neutralinos can be produced highly boosted from a squark or gluino decay. 
By investigating 
the jet substructure, it was shown that a signal 
can be extracted from the large light-jets background without making any 
assumptions on the presence of charged leptons. The neutralino mass was 
determined to ${\mathcal O}(10~{\rm GeV})$ precision. 

In Ref.~\cite{ATL-PHYS-PUB-2009-076}, these methods 
were tested in a full ATLAS simulation. Using the $k_T$ jet 
algorithm with a size parameter of $R=0.7$, the neutralino decay 
was shown to 
produce a single fat-jet when the neutralino transverse momentum exceeded 
a few hundred \GeV. The technique was further tested,
at the Tevatron, in Ref.~\cite{Brooijmans:2010tn} to investigate the possibility of 
reconstructing very light gluinos ($m_{\tilde g}\sim 150$~\GeV) 
decaying into three quarks.

\section{Tools and techniques}
\label{sec:had:tools}


Jet substructure analyses are able to distinguish the fat-jets, which form 
when highly boosted particles decay to quarks or gluons, from the large 
QCD jet background.

A sophisticated set of tools has been developed to try to answer
the following two questions:
Firstly, given a massive jet, is its mass due to the presence of a
decayed massive object (signal) or simply a consequence of the QCD
emissions that always occur within jets produced by light quarks or gluons (background)?
Secondly, assuming a jet does come from a decayed massive object, can
one establish which particles are most likely to have come from that
massive object and which ones are more likely to be due to
initial-state radiation, underlying event (UE) or pileup (PU)? 
This second question is important because the addition of UE/PU particles 
to a jet can severely degrade mass resolution.

Three (somewhat overlapping) sets of methods have been developed to help 
address these questions: identification
of subjets within the candidate jet, dedicated ``grooming'' away of uncorrelated
radiation within a jet and energy flow techniques.

\subsection{Two-body subjet methods}
\label{sec:jet-substr-tools}

Subjet methods are mostly based on the $k_T$~\cite{kt,kt2} and
Cambridge/Aachen (C/A)~\cite{Dokshitzer:1997in,Wobisch:1998wt} jet
clustering algorithms, either directly on jets found with these
algorithms, or on the result of reclustering some other jet's
constituents.
These algorithms sequentially merge (by four-vector addition) the pair of particles that are
closest in some distance measure $d_{ij}$,
\begin{equation}
  \label{eq:dij}
  d_{ij} = \min(p_{Ti}^{2n}, p_{Tj}^{2n}) \, \frac{\Delta R_{ij}^2}{R^2}\,,
  \qquad
  \begin{cases}
    k_T: \quad & $n=1$\,,\\
    \mathrm{C/A}:  \quad & $n=0$\,,
  \end{cases}
\end{equation}
unless there is a distance $d_{iB} = p_{Ti}^{2n}$ which is smaller
than all $d_{ij}$, in which case particle $i$ is called a jet and the
clustering proceeds with the remaining particles in the event\footnote{The parameter 
$\Delta R_{ij}$ is the (angular) distance between constituents $i$ and $j$ in $y$-$\phi$ space, 
where $y$ is the rapidity and $\phi$ is the azimuthal angle of the constituents transverse to the beam direction. 
The parameter $R$ controls the size of the jets in $y$-$\phi$ coordinates and can be roughly referred 
to as the jet radius (even though these jets are not usually circular).  In particular, the 
criteria above ensure that particles separated by $\Delta R > R$ at a given clustering 
stage cannot be combined and that a particle can only be promoted to a jet if there are 
no other particles within $\Delta R < R$.}.
Many jet substructure methods undo one or more steps of the clustering
so as to identify subjets that correspond (approximately) to the
individual decay products of the massive object\footnote{Though the anti-$k_T$ algorithm is 
formulated similarly to $k_T$ and C/A, simply with $n=-1$ in Eqn.~\ref{eq:dij}, where $n$ governs the relative power of the energy versus geometrical scales, 
its intrinsic hierarchy is in general unsuitable for direct substructure studies. 
This is because when it merges a softer subjet with a main hard subjet, the merging
often takes place across a multitude of clustering steps. For the $k_T$
and C/A algorithms, such a merging happens in a single step.}. 
Utilities for clustering and for studying the 
clustering history are available in {\tt FastJet}~\cite{fastjet,fastjetURL}.
Additional analysis tools are supplied by {\tt SpartyJet}~\cite{SpartyJet,spartyjetURL}.

In the $k_T$ algorithm, the final step in the
clustering of a jet usually corresponds to the merging of
the two decay products of the massive object.
This was exploited in an early study by Seymour~\cite{Seymour:1993mx}
of boosted $W$ boson decays, which involved undoing the last stage of a ($R=1$)
$k_T$-jet's clustering to obtain two subjets.  In one analysis, this was
followed by angular cuts
on the separation between those two subjets, to reduce backgrounds.  In particular,
QCD jets with masses near $m_W$ tend to acquire much of their mass from 
relatively soft parton branchings at wide angles.  In another analysis in \cite{Seymour:1993mx} 
the subjet separation was used
to set a smaller radius
for a more refined, second stage of $k_T$ clustering on the $W$-jet constituents.  Particles from the leading two refined
jets were used to reconstruct the $W$ boson, 
thereby ignoring wider-angle radiation, usually dominated by 
UE/PU.\footnote{The angular-ordering property of QCD tells us that to catch the QCD
radiation from the decay products of a colour-singlet such as the $W$ boson, the subjets need only extend out as
far as the subjet angular separation.  Any particles found beyond this tend to be uncorrelated.}

In a procedure dubbed ``YSplitter,'' Butterworth, Cox and 
Forshaw~\cite{Butterworth:2002tt} (see also~\cite{butterworth,Butterworth:2007ke,cscnote}) 
also used the $k_T$ algorithm and simply cut on the value of the $d_{ij}$ distance (equivalently, the $k_T$ scale) 
in the final merging.  QCD backgrounds tend toward small values (depending
in detail on the jet definition and $p_T$ scale); whereas $W$-jets tend toward values correlated with 
$m_W$.\footnote{This strategy is similar to Seymour's subjet angular cut method, if we restrict our
attention to jets with mass near $m_W$.  In the approximation of massless subjets, $k_T$ and $\Delta R$ are strictly
related for fixed originator mass.}$^,$\footnote{The possible degradation of mass resolution due to UE/PU is not
addressed by this procedure, but YSplitter can readily be combined with dedicated grooming approaches
if necessary.}

Butterworth, Davison, Rubin and Salam (BDRS)~\cite{Butterworth:2008iy} pointed out that the C/A
algorithm, which repeatedly clusters the closest pair of particles in angle, produces a 
much more ``angular-ordering aware'' organisation of jet substructure.  This observation
served as the basis for a method that combines both improvement of mass resolution and reduction 
of internal phase space in a relatively scale-free manner.  It is particularly well-suited 
for performing unbiased searches for undiscovered boosted 2-body resonances, such as the Higgs, since 
QCD-initiated jets processed by this method produce a relatively featureless mass spectrum.

In contrast to the $k_T$ algorithm, it is not useful 
just to undo the last stage of C/A clustering: The absence of any
momentum scale in its distance measure means that the last clustering
often involves soft radiation on the edges of the jet and so, is
unrelated to the heavy object's decay.
C/A-based approaches must therefore continue to work backwards through the jet
clustering and stop when the clustering meets some specific hardness
requirement.  BDRS require a substantial ``mass-drop'',
i.e.\ $\max(m_i,m_j)/m_{ij}$ significantly below $1$, and symmetry between the momentum fractions of
the two subjets, expressed as a cut on:\footnote{Alternate hardness measures are also possible,
such as the softer subjet's $p_T$ divided by the total jet $p_T$, the angular separation
between the subjets and Jade-type distance measures, e.g. \ $d_{ij}^{(Jade)} =
p_{Ti}p_{Tj} \Delta R_{ij}^2$.}$^,$\footnote{The mass-drop criterion identifies a localised region 
within the fat-jet that looks like two distinct cores of energy.  The asymmetry cut 
essentially serves the same purpose as the $k_T$ scale cut in \mbox{YSplitter}, to eliminate energy-sharing 
configurations that look more QCD-like.  But, as phrased here, the cut no longer refers 
to an absolute mass scale.  
Declustering can continue indefinitely, down to arbitrarily small
$\Delta R$ and arbitrarily small masses, until suitable substructure is 
identified.} 
\begin{displaymath}
d_{ij}^{(n=1)}R^2/m_{ij}^2\simeq \min(p_{Ti},p_{Tj})/\max(p_{Ti},p_{Tj}).
\end{displaymath}
 These criteria are controlled by two dimensionless thresholds, $\mu$ 
and $y_{\rm cut}$, respectively.

While wide-angle UE/PU radiation is actively removed by the mass-drop 
procedure, this removal is not sufficient in the moderately-boosted
regime, $\Delta R \simeq 1$, studied by BDRS.  
The subjets obtained in this way may still be quite large and 
contaminated.\footnote{UE and PU contamination to the jet mass scales as $R^4$
\cite{Dasgupta:2007wa}.}  To refine the subjets further, BDRS apply a ``filtering''
procedure, close in spirit to the reclustering method of Seymour~\cite{Seymour:1993mx}.
The constituents of the two subjets are reclustered with C/A, using 
$R = {\rm min}(0.3, \Delta R_{\rm subjets}/2)$.  The three hardest subjets are
taken, facilitating the capture of possible gluon radiation in the heavy particle decay,
while still eliminating much of the UE/PU. The procedure was adapted for top-taggers in
\cite{Plehn:2009rk,Plehn:2010st} and found to be beneficial in normal
dijet studies too \cite{Cacciari:2008gd}. A discussion of its
optimisation for two-body decays is given in Ref.~\cite{Rubin:2010fc}.

\subsection{Three-body subjet methods: top-taggers}

A variety of three-body subjet methods have been developed, building on the two-body methods.
These have mainly been tailored for boosted tops, but have also been adapted for generic
heavy particle searches.

One of the simplest top-taggers is an extension of \mbox{YSplitter} by 
Brooijmans~\cite{brooijmans}.  Further substructure is revealed by repeated
$k_T$ declustering and reading off the $k_T$ scales of the next-to-last (and next-to-next-to-last) 
clusterings.  Top-jets can be discriminated from QCD by placing cuts in the multidimensional 
space of jet mass and $k_T$ scales, or by using a likelihood ratio built in this space~\cite{brooijmans2}.
In subsequent sections, we refer to this as the ``ATLAS'' tagger.

Thaler and Wang~\cite{Thaler:2008ju} utilise a similar approach.  A jet is reclustered with $k_T$,
until, depending on the analysis, exactly two or three subjets are formed. 
Internal kinematic variables in addition to $k_T$ scales are utilised.  For example, in the three-subjet analysis, a
$W$ boson candidate is identified by forming the minimum pairwise mass between subjets and a minimum
cut is placed on its mass.  
Relative energy sharings between the subjets are also studied.

The ``Hopkins'' top-tagger of Kaplan, Rehermann, Schwartz and 
Tweedie~\cite{Kaplan:2008ie,hopkinstaggerURL} is a descendant 
of the two-body approach of BDRS.  The original version is specialised for a perfect 
$\Delta\eta\times\Delta\phi = 0.1\times0.1$ calorimeter, as a straw-man detector.  
Quantities $|\Delta\eta|+|\Delta\phi|$ and 
${\rm min}(p_{T1},p_{T2})/p_{T{\rm jet}}$ are used to categorise clusterings, rather than 
relative mass drop and pairwise energy asymmetry.  Thresholds 
are set by parameters $\delta_r$ and $\delta_p$, respectively.   When an interesting declustering is found, the two 
subjets are used as ``jets'' for a secondary stage of two-body substructure searches.  
The original jet is a good top candidate if at least one of these secondary 
declusterings succeeds, so that there are three or four final subjets.\footnote{At all stages, 
relative $p_T$ is measured with respect to the $p_T$ of the entire original jet, so that the 
threshold is not sensitive to internal energy distributions within the jet.}  Kinematic cuts are then applied
to these subjets (without filtering): The summed subjet mass should lie near $m_t$, there should be a 
subjet pair that reconstructs near $m_W$ and the reconstructed $W$ boson helicity angle should not
be too shallow.\footnote{The $W$ boson mass constraint ensures that at least one subjet pairing achieves sizable
invariant mass.  The helicity angle cut then removes QCD-like configurations where one of the other subjet pairings
produces a small mass.}

The Hopkins tagger has been modified by CMS~\cite{cmsnote3,cmsnote2,Rappoccio:2009cr} in two ways.
First, as the actual LHC detectors have better resolution than the $0.1\times0.1$ grid model, declustering
uses $\Delta R$ to determine adjacency and the parameter $\delta_r$ shrinks with 
$p_T$:  $\delta_r = 0.4 - 0.0004\cdot p_{T{\rm jet}}$.  Second, the $W$ boson mass window and helicity angle cuts are replaced
by a single cut on the minimum pairwise subjet mass (excluding any fourth subjet), as in 
Thaler and Wang\footnote{This achieves similar performance to the original cuts, which are essentially forcing
the same type of constraint on internal kinematics.}.

A method more closely tied to BDRS and building on the
top-Higgsstrahlung study of~\cite{Plehn:2009rk}, appears in a paper by 
Plehn, Spannowsky, Takeuchi and Zerwas~\cite{Plehn:2010st}, the
Heidelberg-Eugene-Paris (HEP) top-tagger.  A fat-jet is declustered using 
a fractional mass-drop criterion, with no asymmetry requirement.  Subjets-within-subjets are searched for
indefinitely, until subjets with masses below 30 GeV are encountered.
As the boosts in this study are modest and the $\Delta R$ scales are quite large, multibody filtering 
is also applied.
For every set of three subjets, one reclusters the constituents with
C/A, $R = {\rm min}(0.3,(\Delta R_{jk}/2))$ ($j,k$ run over the
three subjet indices) and takes the invariant mass of the 5 hardest
resulting filtered subjets as a filtered mass.
The set of three initial subjets that gives a filtered mass closest to
the top mass is retained as the sole top candidate. 
The set of filtered constituent particles is then reclustered yet again
to yield exactly three subjets and
non-trivial cuts are placed in the 
two-dimensional subspace of $(m_{23}/m_{123},{\rm arctan}(m_{13}/m_{12}))$, where the numbers 
label the subjets in descending $p_T$.  For tops, one of the pairings will have a relative 
mass $m_{ij}/m_{123} \simeq m_W/m_t$ and the cuts are designed to capture this region.  
All cuts for the present tagger have been designed to be free of explicit mass scales by normalising 
with respect to the jet mass.

Tagging of boosted 3-body decays can also be applied to search for
as-yet undiscovered particles.  Of particular note are the
R-parity-violating neutralino taggers of Butterworth, Ellis, Raklev
and Salam~\cite{Butterworth:2009qa}, which generalize the idea of
3-body tagging beyond the special kinematic situation of the top.  The
paper explores two tagging methods. One is similar to the 
YSplitter top-tag in Ref.~\cite{brooijmans,brooijmans2} but with cuts on 
dimensionless ratios such as $d_{12}/m^2$. 
The other is C/A based and searches the entire clustering history of a
jet, paying attention only to clusterings that are not too locally
$p_T$-asymmetric (using a single asymmetry parameter $z_{\rm min}$)
and recording the associated Jade-distances, $p_{T1}p_{T2}\Delta
R_{12}^2$.
The clustering with the largest Jade-distance defines the neutralino
candidate and to ensure 3-body kinematics, a cut is placed on the ratio
of the masses of the subjets with second-largest and largest Jade
distances.
An adaptation of this method has been used in an unpublished C/A-based
top-tagger by Salam~\cite{CMtaggerURL}, which takes the subjet with
the largest Jade-type distance as the top candidate. Then in the
top candidate rest frame, the top constituents are reclustered into
exactly three jets using the $e^+e^-$ style $k_T$ algorithm
(see~\cite{Salam:2009jx} for a description).  The hardest subjet (in
absolute energy) is assumed to be the $b$-jet and the other two
constitute the $W$ boson decay products.  Three-body kinematics can then be
constrained as desired.


\subsection{Jet grooming methods}
\label{sec:had:tools:groom}

Jet grooming - elimination of uncorrelated UE/PU radiation from a target jet - is useful irrespective of the specific
boosted particle search and can even be applied for slow-moving heavy particles that decay to well-separated jets.  
While many of the methods discussed above incorporate some form of grooming, notably Seymour
and BDRS (filtering), two other methods specifically dedicated to grooming have been developed.

Pruning was introduced by Ellis, Vermilion and Walsh \cite{Ellis:2009su,Ellis:2009me}.  
The idea is to take a jet of interest and then to recluster it using a vetoed 
sequential clustering algorithm.  Clustering nominally proceeds as usual, but it is 
vetoed if 1) the particles
are too far away in $\Delta R$, {\it and} 2) the energy sharing, defined by min$(p_{T1},p_{T2})/p_{T(1+2)}$, is too asymmetric.
If both criteria are met, the softer of the two particles is thrown away and all
$d_{ij}$'s and $d_{iB}$'s  are recalculated.  The $\Delta R$ and energy-sharing thresholds are set by adjusting the two 
parameters $D_{\rm cut}$ and $z_{\rm cut}$, 
respectively.\footnote{Jets built in this manner will be stripped of soft wide-angle radiation from the bottom-up.  As an added
bonus, C/A substructure organisation is now rendered trivial, much like $k_T$.  Interesting 
substructures within pruned C/A jets tend to live in the final clustering stages.}

Trimming is a technique that ignores regions within a jet that fall 
below a minimum $p_T$ threshold.  It was introduced by Krohn, Thaler and Wang in~\cite{Krohn:2009th}.  
(Similar ideas also appear
in~\cite{Butterworth:2008iy,Seymour:1993mx}.) Trimming reclusters the
jets' constituents with a radius $R_\text{sub}$ and then accepts only the subjets that have
  $p_{T,\text{sub}} > f_\text{cut}$, where $f_\text{cut}$ is taken
  proportional either to the jet's $p_T$ or to the event's total $H_T$.  The small jet 
radius and energy threshold are the only parameters.

While different grooming methods have the same goal, it may be 
possible to combine approaches for greater effectiveness.  For a first study along these lines,
see~\cite{Soper:2010xk}.

\subsection{Event-shape and energy-flow methods}
\label{sec:other-tools}

Top decays often feature a triangular structure, 
transverse to the boosted top-quark axis.
Event-shape type measures (widely used at $e^+e^-$ colliders) can be
applied to the jet constituents to help establish whether such
a triangular structure is present. 
Refs.~\cite{Thaler:2008ju,Almeida:2008yp} both proposed planar-flow
type observables, for which one establishes the eigenvalues of a
matrix such as
\begin{equation}
  \label{eq:planar-flow}
  I^{kl} = \frac{1}{m_J} \sum_{i \in \mathrm{jet}} \frac{p_{i}^{\perp k}\, p_{i}^{\perp l}}{E_{i}}\,,
\end{equation}
where $p_{i}^{\perp k}$ is the $k^\mathrm{th}$ component of $p_{i}$
transverse to the jet axis and $E_{i}$ is its energy. The matrix has
two eigenvalues, $\lambda_1 \ge \lambda_2$, and the second eigenvalue provides a
measure of the jet planarity, for example through the combination
$4\lambda_1\lambda_2/(\lambda_1+\lambda_2)^2$.

Another method of making use of the energy flow \cite{Almeida:2010pa}
involves the construction of energy-flow ``templates''. These take the
energy flow, discretised in $\theta$ and $\phi$ (templates), for each
possible orientation of top-decay products (possibly with cuts to
limit backgrounds). 
Then for a given jet in an event, the method finds the template that
provides the best match event to that jet's energy flow pattern, with
a measure of the match quality that involves Gaussians of the
difference between actual and template energy flows.

Two other energy-flow/event-shape type methods introduced recently
aim to isolate signals based on the \emph{absence} of energy
flow~\cite{Chen:2010wk,Falkowski:2010hi}: They both involve the (same)
case of colour-neutral particles with $p_T \gg m$ that decay to two
coloured particles $i$ and $j$. Because of the colourless nature of
the parent, the emissions from the coloured particles are highly
collimated within an angle $\Delta R_{ij}$ (once again, due to angular
ordering). In contrast QCD backgrounds involve emission on all angular
scales. Vetoing on energy-flow and/or subjets outside $\Delta R_{ij}$
therefore allows a significant reduction in background while
retaining much of the signal.

In this context it is worth commenting also on the use of
colour-structure dependence of energy flow in the non-boosted limit, to help distinguish signals of colour-neutral
heavy-object decays (two jets colour connected to each other) from
backgrounds (two jets, each colour connected to the beams) through an
observable named ``pull''~\cite{Gallicchio:2010sw}.


\section{Experimental status of jet substructure}
\label{sec:had:exp}

The previous BOOST workshops have highlighted the importance of understanding jet substructure and have spurred numerous groups to use existing data sets to perform studies that address some of the key uncertainties raised in these previous meetings.  This section provides a brief review of pioneering studies of jet substructure using data collected at the DESY HERA $ep$ Collider and the Fermilab Tevatron Collider, as well as the recent work reported for the first time at BOOST2010.

\subsection{Jet substructure measurements performed at HERA}

One of the earliest studies looked at the mean number of subjets in a recoil 
jet produced in the photoproduction of jets in $ep$ 
collisions \cite{ZEUS:subjet2003}.  
From jets produced at large angles to the proton beam and with transverse 
energies $E_T > 17$~GeV, the average number of subjets in the recoil jet 
was used to measure the strong coupling constant and to confirm the general 
picture of QCD radiation within the perturbative parton shower believed 
to be responsible for the jet.  
A subsequent study \cite{ZEUS:subjet2004}\ employed a  sample of jets 
produced through photoproduction and deep-inelastic scattering to study
the kinematics of jet production as well as the distribution of energy flow 
within the jet.  
The data are well-described by QCD calculations as implemented in {\sc PYTHIA} 
and an extraction of the strong coupling constant was made. 
Most recently, the ZEUS collaboration reported a study of  subjets in jets 
produced in neutral-current deep-inelastic scattering \cite{ZEUS:subjet2009}.   
The jets were clustered with the $k_T$ algorithm and the subjet structure was 
obtained in the laboratory frame, by running a variant of the exclusive $k_T$ cluster 
algorithm\footnote{$k_T$-style clustering continues running on the jet constituents, without using $d_{iB}$'s, until all of the $d_{ij}$'s exceed
$y_{\rm cut}p_{T \rm jet}^2$.  The final set of clustered particles are the subjets.} with a 
$y_{\rm cut} = 0.05$\ for jets with $E_T > 14$~GeV and pseudorapidity from -1 to 2.5. The
dimensionless parameter $y_{\rm cut} $ is related to the $k_T$ distance metric through the following 
formula:
\begin{equation}
d_{ij} = \min(p^2_{T,i},p^2_{T,j}) \frac{ \Delta R^2}{R^2} > y_{\rm cut} p^2_T\,,
\end{equation} 
where $R$ is the resolution parameter of the $k_T$ algorithm and $p_{T,i}$, $p_{T,j}$ and $p_T$ denote the transverse momentum of the two subjets and of the parent jet, respectively.

Focusing on the kinematic distributions of those jets with exactly two 
subjets, the study found that QCD predictions were in good agreement with 
the data, again confirming the general picture of QCD radiation in the 
showering process.

\subsection{D0 jet substructure measurements}

D0 studied the $k_T$ subjet multiplicity for central ($|\eta| < 0.5$) jets 
reconstructed with the $k_T$ algorithm
with $R=0.5$\ and 55 GeV$ < p_T < 100$~GeV in data collected during 
Run~I of the Fermilab Tevatron Collider at $\sqrt{s} = 0.63$\ and 1.8 TeV \cite{DZero:subjet2002}.  
The analysis selects subjets based on $y_{\rm cut} = 10^{-3}$. The choice to a minimum 
subjet $p_T$\ of approximately 3\%\ of the total jet $p_T$.

The subjet $p_T$\ distribution shows that jets are composed of a soft and a 
hard component.  
The soft component has a threshold at 1.75~GeV set by the value of 
$y_{\rm cut}$\ and the minimum jet $p_T$, whereas the 
hard component peaks at 55 GeV driven by single-subjet jets.  
Exploiting the two centre-of-mass energies
and taking the fraction of gluon jets at each of these from simulation, the 
subjet multiplicity for quark and gluon jets is extracted from the data.  
After correcting for subjets originating from showering in 
the calorimeter, as well as other small effects, the ratio of the 
average number of {\em extra} 
subjets (i.e. average minus one)
in gluon relative to quark-originated jets is measured to be 
$1.84 \pm 0.15 ({\rm stat.}) \pm 0.20 ({\rm syst.})$, confirming 
that gluon jets radiate more than quark jets.

\subsection{CDF jet substructure measurements}

CDF performed an early Run II measurement of jet substructure using 
0.17 fb$^{-1}$\ of data collected at $\sqrt{s} = 1.96$~TeV.  Measurements were 
carried out on jets with $p_T$\ up to 380 GeV, with the key variable being the 
average fraction of jet transverse momentum that lies inside a cone of radius $r$\ 
concentric to the jet cone, as a function of $r$\ \cite{CDF:subjet2005}.

These measurements showed that {\sc PYTHIA}~\cite{pythia} (v. 6.203) calculations with Tune A settings 
provided a reasonable description of the observed data. The {\sc herwig} 6.4~\cite{herwig} MC 
calculations also gave a reasonable description of the measured jet shapes, 
but tended to produce jets that were too narrow at low jet $p_T$ values.

CDF presented new results at the BOOST2010 meeting of measurements of jet mass, 
angularity and planar flow for jets with $p_T > 400$~GeV from a sample of 
5.95 fb$^{-1}$\ \cite{CDF:subjet2010a}.  
The measured distributions were compared with analytical expressions from NLO 
QCD calculations, as well as {\sc PYTHIA} 6.1.4 predictions incorporating full detector 
simulation.  
The theory predictions for jet mass were in good agreement with the data, 
whereas the angularity and planar flow predictions by {\sc PYTHIA} showed disagreement 
in detail (primarily at low angularity and low planar flow).   
Subsequent to the meeting, CDF presented results of a search for boosted top 
quarks in this sample, setting a preliminary upper limit of 54 fb on Standard Model top 
quark production cross section for top quarks with $p_T > 400$~GeV at 95\%\ confidence level \cite{CDF:subjet2010b}.

\section{Benchmark samples}
\label{sec:had:samples}
ç

Over the years the community involved in studies of boosted objects has 
grown considerably. A large number of different tools exists and
it is often hard to gauge their relative performance from published results.
Moreover, as we will show in section~\ref{sec:had:robust}, the choice
of Monte Carlo tools used to model the jets can have a pronounced impact 
on the results. We have therefore generated a number of common samples
to provide a benchmark for the performance comparisons in the following sections.
 
We simulated LHC proton-proton collisions at a centre-of-mass energy of 7 TeV. 
The samples consist of QCD dijet events, representing the most important 
background for many searches, and the Standard Model $t\bar{t}$ events 
as signal, serving as a typical example of heavy, boosted particle production. 
Both samples were produced with {\sc HERWIG} 6.510~\cite{herwig}.  All samples 
are divided in equally-sized sub-samples with parton \pt ranges from 200-300 
\gev, 300-400 \gev, \ldots, 1.5--1.6 \tev, thus covering the full range from 
topologies with moderate boost to extremely energetic events. We generated
10.000 events for each parton \pt bin. Combining all samples yields an 
approximately flat $p_T$ distribution. 

The generated events include a description of the underlying event (UE). 
{\sc HERWIG} is used in conjunction with {\sc JIMMY}~\cite{Butterworth:1996zw} that takes care of the 
underlying event generation. For this study we rely on a tune from 
ATLAS~\cite{ATL-PHYS-PUB-2010-002}~\footnote{The parameters of this tune are set as such: 
{\sc PDF = MRST2007lomod}, {\sc PTJIM}~$=3.6 \times (\sqrt{s}/1800~{\rm GeV})^{0.274}$, {\sc JMRAD(73)} = 2.2, {\sc JMRAD(91)} = 2.2 and {\sc PRSOF} = 0.0 (i.e. {\sc HERWIG}'s internal soft UE turned off).}.

The subleading terms in parton generators are often constrained by tuning. Extensive tuning has been performed at LEP, mostly constraining quark jets. However, little tuning of substructure observables has been performed at hadron colliders, where, relative to LEP various new elements arise: there are many more gluon jets, there are colour connections with the initial state, and there is the question of how to handle recoil from hard emissions in a context where the partonic centre-of-mass energy is no longer fixed. We therefore expect that the description of jet substructure in different MC tools may differ significantly. 

For comparison we generated identical samples with {\sc PYTHIA} 6.4~\cite{pythia}. A number of
tunes for the UE description were considered, that we 
will label as DW, DWT and Perugia0. The parton shower model of the DW and 
DWT samples is $ Q^2$-ordered. Both yield identical results for the underlying 
event at the Tevatron. However, the two tunes extrapolate differently to the 
LHC, where DWT leads to a
more active underlying event~\footnote{The value 
of the {\sc PYTHIA} parameter {\sc PARP(90)} governing 
the energy dependence is set to 0.16, the value used by ATLAS, in DWT, while 
the Tevatron tune A value of 0.25 is chosen in tune DW. For a detailed 
description, the reader is referred to Ref.~\cite{Albrow:2006rt}.}. 

The Perugia tune~\cite{Skands:2010ak} uses a $p_T$-ordered parton shower. 
To disentangle the impact of the parton shower and that of the underlying event,
we generated an additional set of {\sc PYTHIA} samples with the UE 
generation switched off. Samples with UE switched off were also produced with {\sc HERWIG}. 

The different generators and tunes used should give a rough measure of the uncertainties in the parton shower and underlying event modelling at the start of data taking at the LHC. We note that, with more refined tunes to LHC data, these uncertainties are expected to decrease considerably over the coming years.

We propose using these samples as a benchmark for MC studies investigating the 
prospects of searches using boosted objects. The samples are publicly available
for future work at:

\begin{itemize}
\item http://www.lpthe.jussieu.fr/$_{\widetilde{~}}$salam/projects/boost2010-events/
\item http://tev4.phys.washington.edu/TeraScale/boost2010/
\end{itemize}

We also agreed on a standard definition of the primary jet reconstruction.
For each event, jets are clustered with the anti-$k_T$ algorithm with an 
$R$-parameter of 1.0. 
As input to the jet clustering, all stable particles with $|\eta| < 5.0$ 
except neutrinos and muons are used. In order to exclude soft additional
jets that do not originate from hadronic top quark decay,
at most 2 jets per event with $\pt > 200\GeV$ are considered.

\section{Impact of jet grooming tools}
\label{sec:had:comp_groom}

Having described many jet substructure tools in section \ref{sec:had:tools}, we now consider how they perform in reconstructing  the hadronic decays of heavy particles.  We begin with three grooming tools: pruning, trimming, and filtering.  Although initially formulated in different contexts, in practice they rely on the same phenomena: Contamination from the underlying event and pile-up will have characteristically lower energy than the core(s) of a high-$\pt$ jet, and most of the energy of the ``uncontaminated'' jet is located in some small number of small regions. Each of the grooming techniques differs in the way this broad idea is implemented and as a result, differences in performance may be expected.

For the purpose of this report, we consider their performance in identifying top jets on the benchmark samples described in Section~\ref{sec:had:samples}. For simplicity we only consider two narrow $\pt$ ranges: 300--400 \gev and 500--600 \gev.

\begin{figure*}[t]
\begin{center}
\subfigure[dijets, 500--600 \gev]{
\includegraphics[width=0.95\columnwidth]{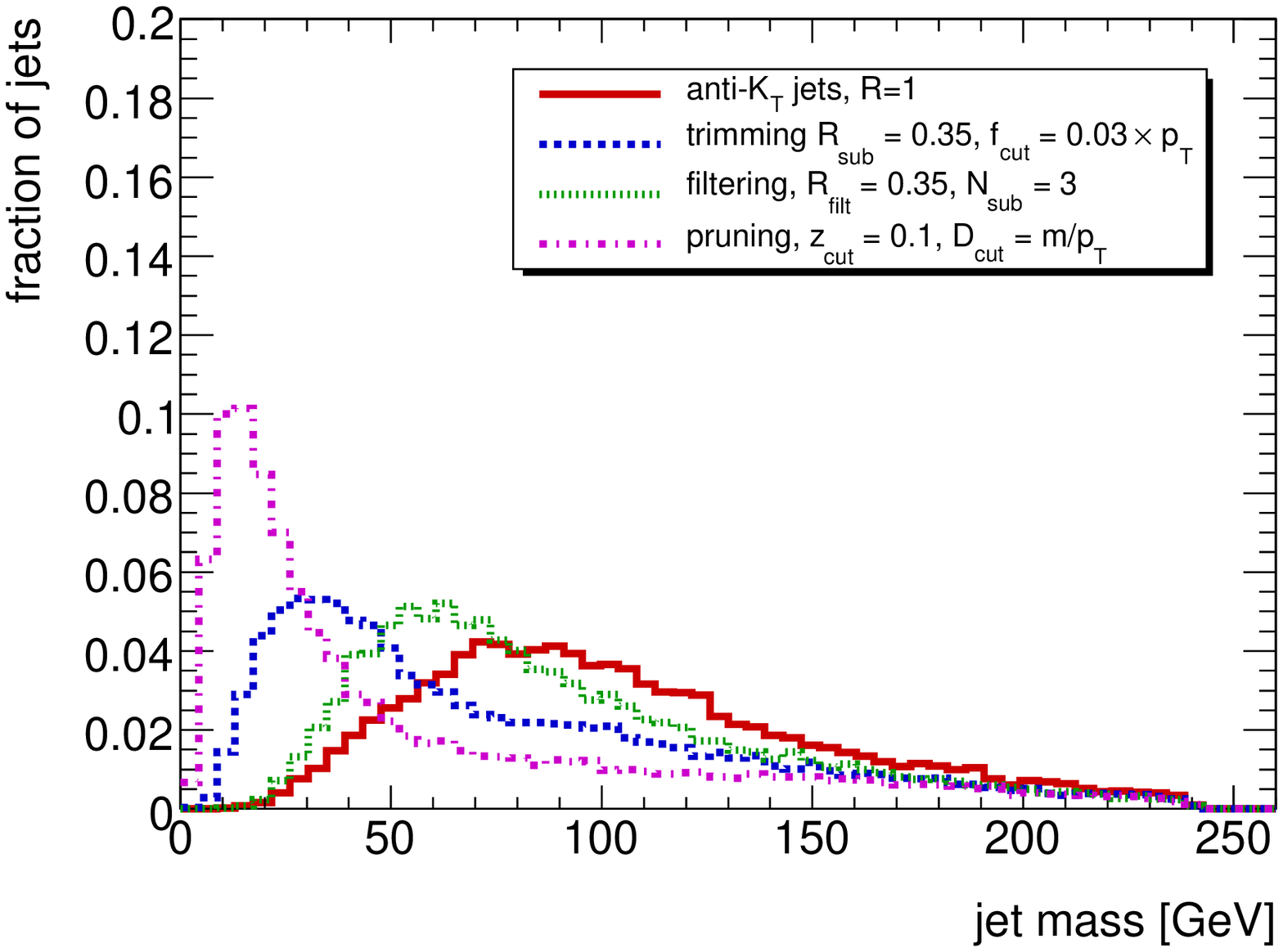}}
\subfigure[$\ttbar$, 500--600 \gev]{
\includegraphics[width=0.95\columnwidth]{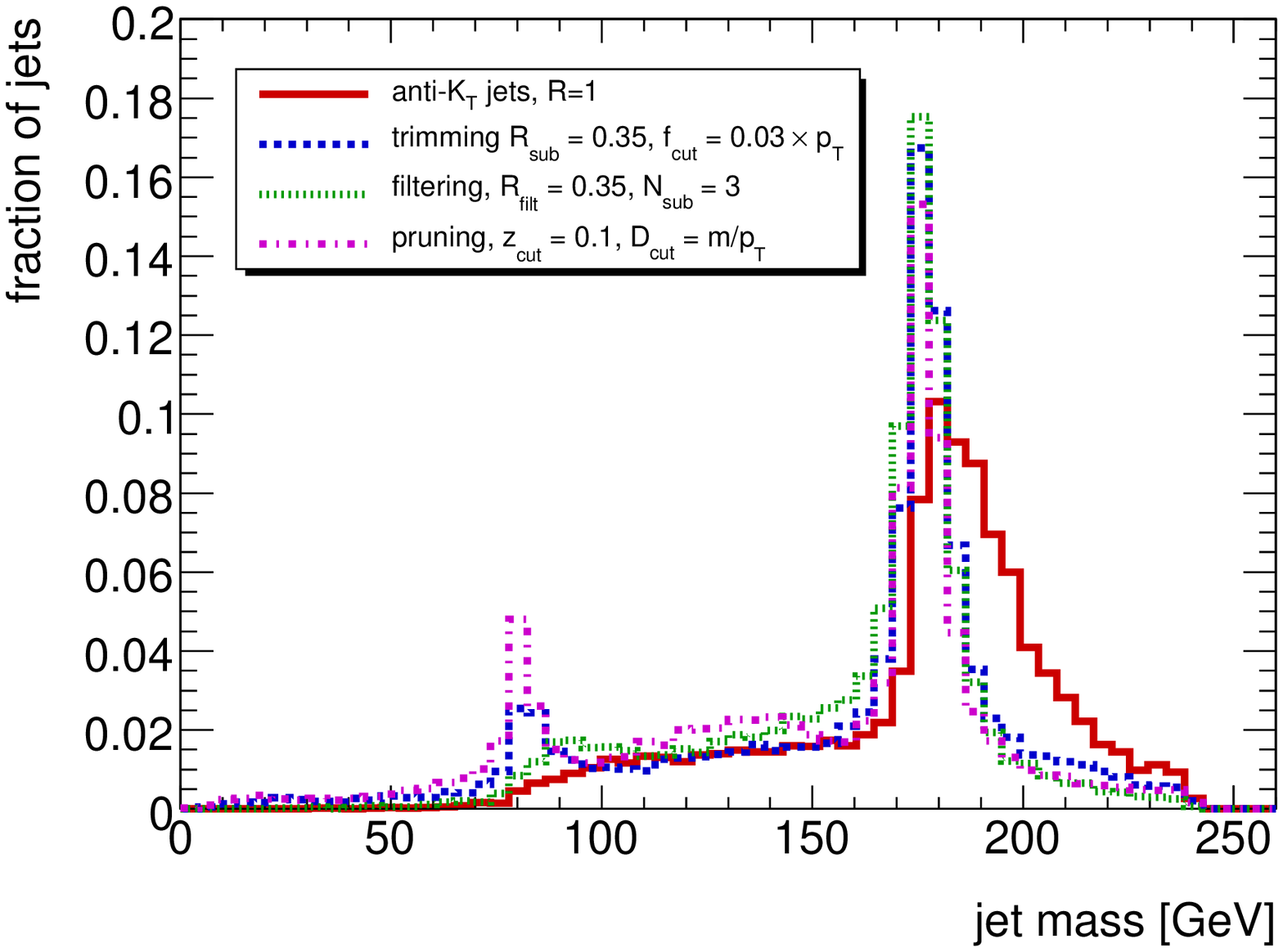}}  
\subfigure[dijets, 300--400 \gev]{
\includegraphics[width=0.95\columnwidth]{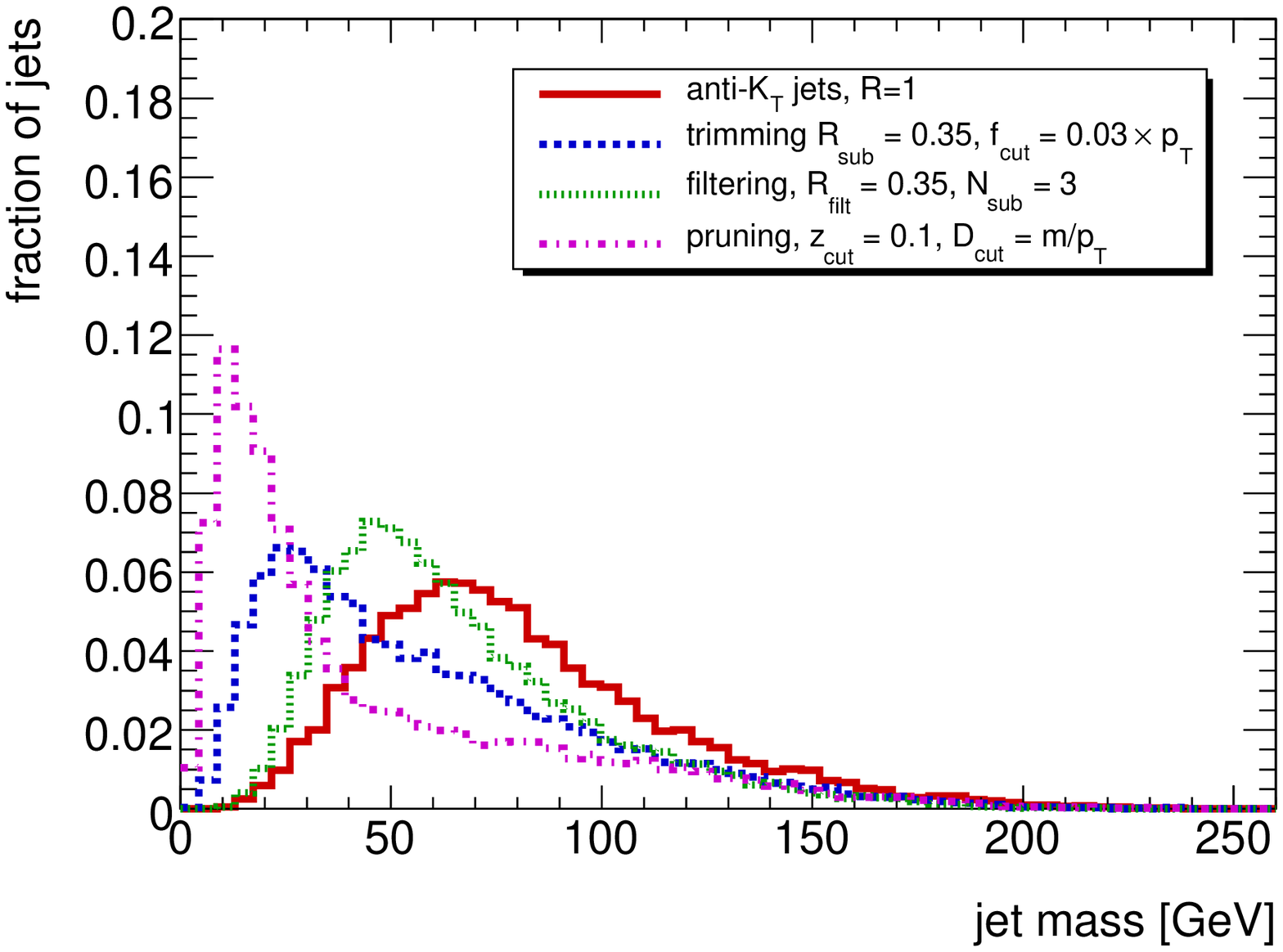}}   
\subfigure[$\ttbar$, 300--400 \gev]{
\includegraphics[width=0.95\columnwidth]{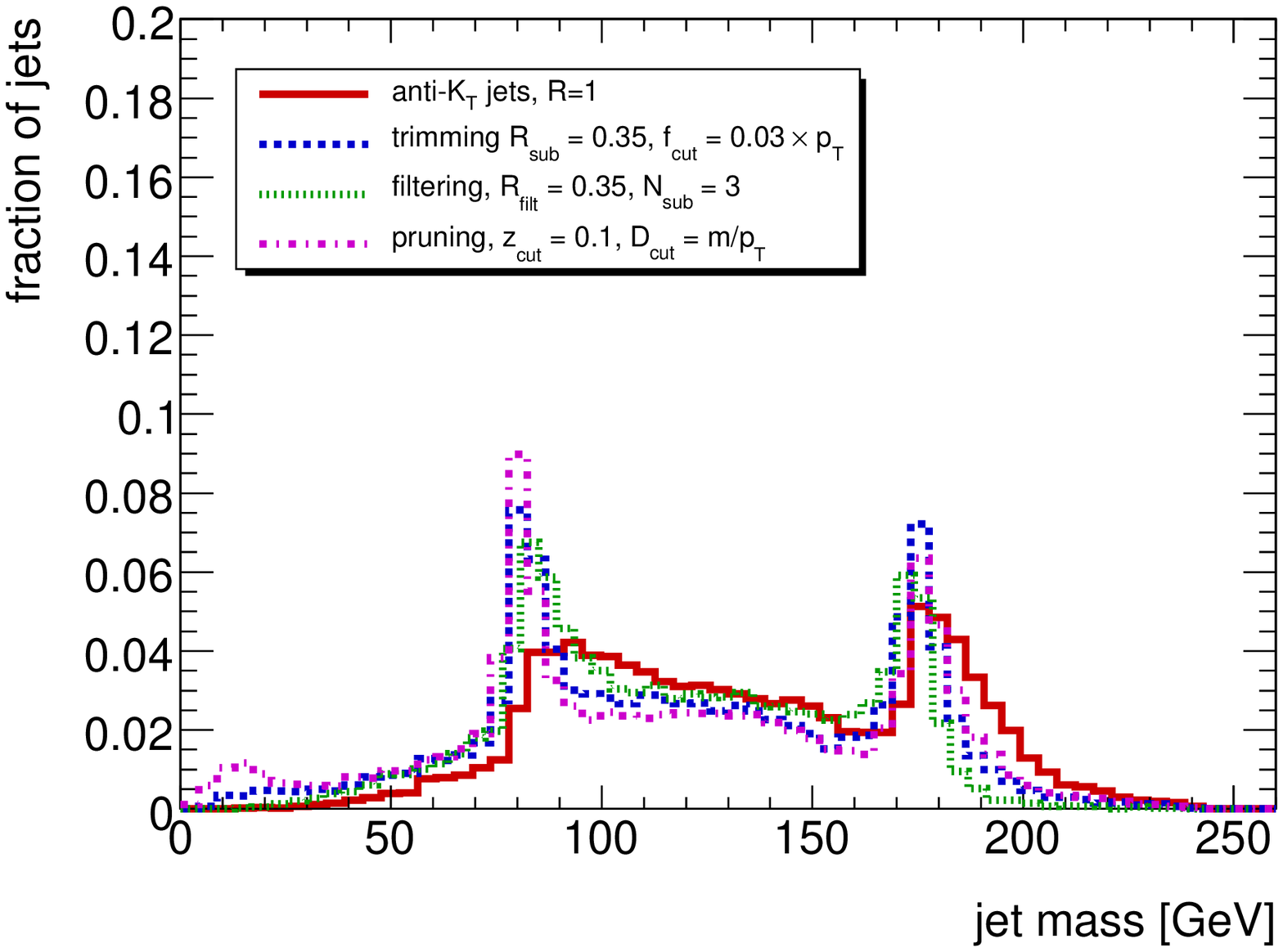}}   
\end{center}
\caption{Jet invariant mass $m_j$ for $\ttbar$ (a,c) and dijet (b,d) events, for three grooming methods.  Each groomed analysis begins with anti-$k_T$ jets with $R = 1.0$.  The solid curve (red in the online version) represents these jets without grooming. The distributions correspond to \ttbar or di-jet quarks or dijet samples with parton-level $\pt$ of 500--600 \gev (a,b) and 300--400 \gev (c,d).} 
\label{fig:had:comp_groom:mjet}
\end{figure*}

For each groomer, the components of anti-$\KT$, $R=1.0$ jets are reclustered with Cambridge/Aachen.  Each groomer then acts on the C/A substructure.  For pruning, which was intended to be used in this manner, we use the ``standard'' parameters from Ref.~\cite{Ellis:2009su}, $\{z_\text{cut} = 0.1, D_\text{cut} = 0.5\times(2m/\pt)\}$.  Trimming was originally proposed for use on QCD jets (while, for example, looking at a dijet mass), so the original parameters are obviously not sensible.  Likewise, filtering has typically been used in the context of further grooming the already identified subjets of a decay.  For trimming and filtering, we have chosen reasonable values based on a superficial exploration of the parameter space, requiring good performance in the higher of the two $\pt$ bins. A careful optimisation of parameters requires a more thorough study. For trimming, we take $\{R_\text{sub} = 0.35, f_\text{cut} = 0.03 \times \pt^\text{jet}\}$. For filtering, we take $\{R_\text{filt} = 0.35, N_\text{subjets} = 3\}$. 

In Fig.~\ref{fig:had:comp_groom:mjet}, we compare the mass distribution for groomed jets with that for ungroomed jets, for QCD dijets (a,c) and for hadronic top decays (b,d).

The most striking difference between the two $\pt$ intervals is the pronounced peak at the $W$ boson mass for 300 $ < \pt < $ 400 \gev sample.  The results in Figures (a) and (c) show that all three grooming techniques affect the shape of the mass distribution of the dijet background. The effect is most clearly visible in the intermediate mass regime, where jet mass is typically dominated by relatively soft radiation around a single, hard core. For the 500 $ < \pt < $ 600 \gev dijet sample the fraction of jets in the mass window from 50--150 \gev is 73 \% for ungroomed jets, and 27 \%, 48 \% and 72 \% for pruned, trimmed and filtered jets, respectively. 

For the high mass tail of the QCD jet mass distribution, large jet masses often come from hard, perturbative emissions that will not be ``groomed away", so grooming diminishes in effectiveness and the differences between the various techniques are less pronounced. For the same sample, the fraction of QCD jets in the mass window from 150--200 \gev is 14.6 \%, 7.8 \%, 9.9 \% and 10.6 \%, respectively.

Turning to the signal distributions of Fig.~\ref{fig:had:comp_groom:mjet} (b) and (d) we find that grooming clearly improves the mass resolution when compared to the raw jet mass. While the resolution of the three grooming methods is similar, the fraction of events in top quark mass peak differs. More aggressive grooming leads to a larger number of signal events that migrate out of the signal peak. For the $\ttbar$ sample with 500 $ < \pt < $ 600 \gev, the fraction of jets with 150 $<m_j < $ 200 \gev is 66 \%, 54 \%, 64 \% and 69 \% for ungroomed, pruned, trimmed and filtered jets, respectively.  

The findings discussed above indicate that while the three grooming techniques have qualitatively similar effects, there are important differences. For our choice of parameters, pruning acts most aggressively on the signal and background, followed by trimming and filtering. These differences can be explained by a more detailed look at the internals of the algorithms.

Filtering is normally used after finding the substructure of the jet and selecting the hardest subjets. In this analysis, however, the number of subjets is fixed to three and even very soft subjets can be included. For this reason filtering is not expected to be as effective in reducing the background in the intermediate mass region.

Trimming, on the other hand, uses a relative $\pt$ threshold to determine which subjets to keep, so soft subjets are discarded.  To a good approximation, low $m/\pt$ QCD jets consist of a single hard core surrounded by soft radiation.  Trimming will keep just radiation within $R_\text{sub}$ of the core; filtering will keep the core as well as two other soft subjets.  In figure~\ref{fig:had:comp_groom:mjet} (a) and (c) we can indeed see that trimming shifts the jet mass distribution further down.  At the same time, the fraction of boosted tops that is found in the top mass interval is slightly reduced.

Like trimming, pruning can strip the jet to a single hard core.  The key difference is that the angular cutoff, $D_\text{cut}$, is adaptive, scaling with the $m/\pt$ of the jet.  This means that at low $m/\pt$, the angular size of the hard subjet(s) kept by pruning gets smaller, making pruning more aggressive. Again we find that this expectation is confirmed by the strong reduction of QCD jets in the intermediate mass regime, and the more pronounced migration of signal events from the top quark mass peak to lower mass.

We conclude that all three grooming methods lead to a significantly improved mass resolution for jets containing a hadronic top quark decay. Aggressive grooming, as implemented in the pruning algorithm, is also very effective in reducing the background population in the intermediate mass regime from 50--150 \gev. For higher masses, dominated by perturbative QCD, grooming techniques are less effective in reducing the QCD background.

\section{Sensitivity of jet substructure to the MC description}
\label{sec:had:robust}

In this section, we study the reliability of Monte Carlo predictions for the substructure of jets. 
To this end, we compare the response of a sub-jet analysis to events generated with several different Monte Carlo tools and UE tunes described in section~\ref{sec:had:samples}. 
In particular, we establish the sensitivity of jet mass and related observables to the parton shower model and to the UE. We also perform a simulation that mimics a number of important detector effects. 
Data collected at the LHC in 2010-2011 should enable a more thorough understanding than we can hope to achieve at this stage. 

\begin{figure*}[t!]
\begin{center}
\mbox{
  \begin{tabular}{ccc}
  \subfigure[jet mass - PS]{\includegraphics[width=0.32\textwidth]{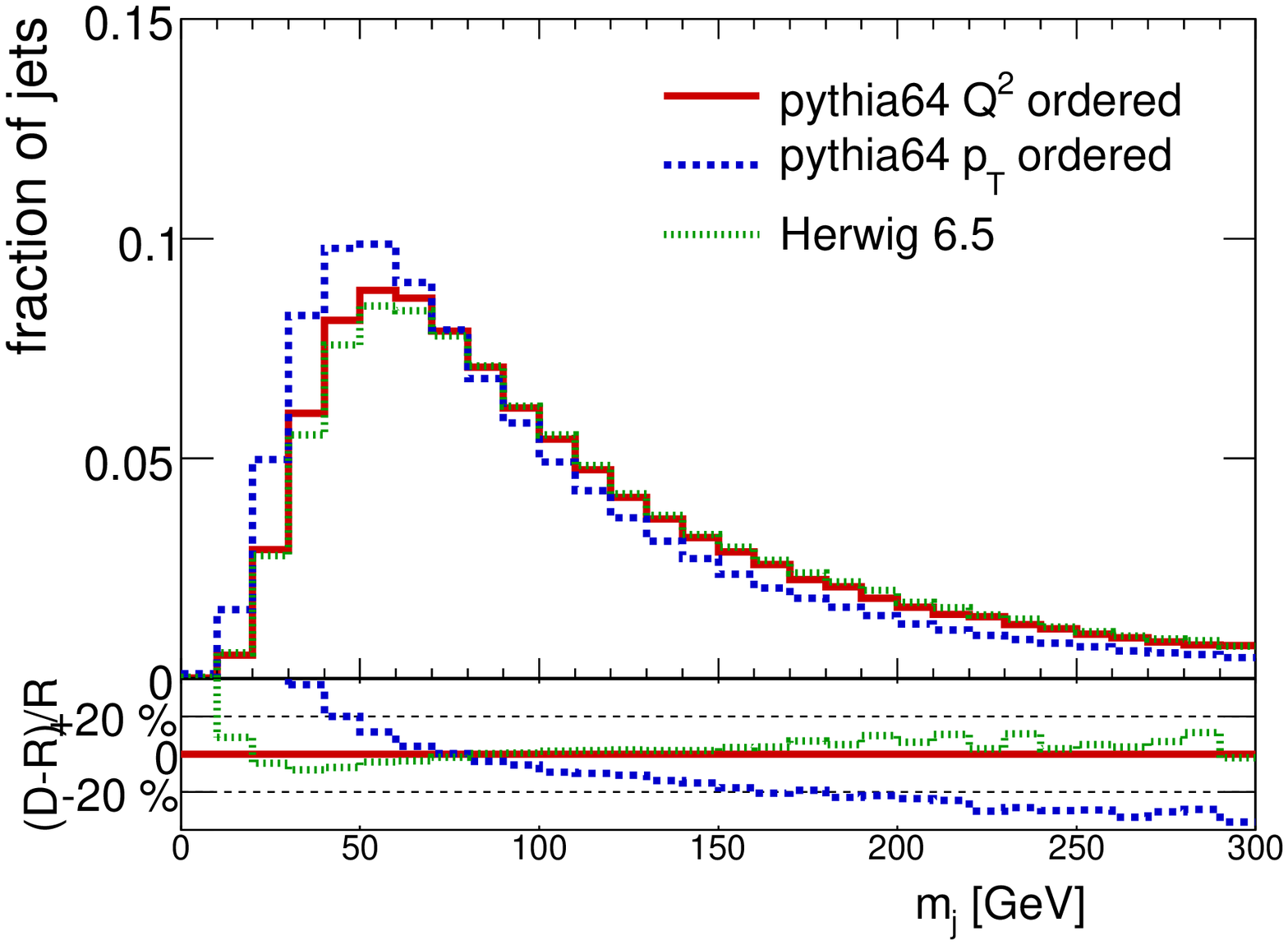}}   
\subfigure[jet mass - UE]{\includegraphics[width=0.32\textwidth]{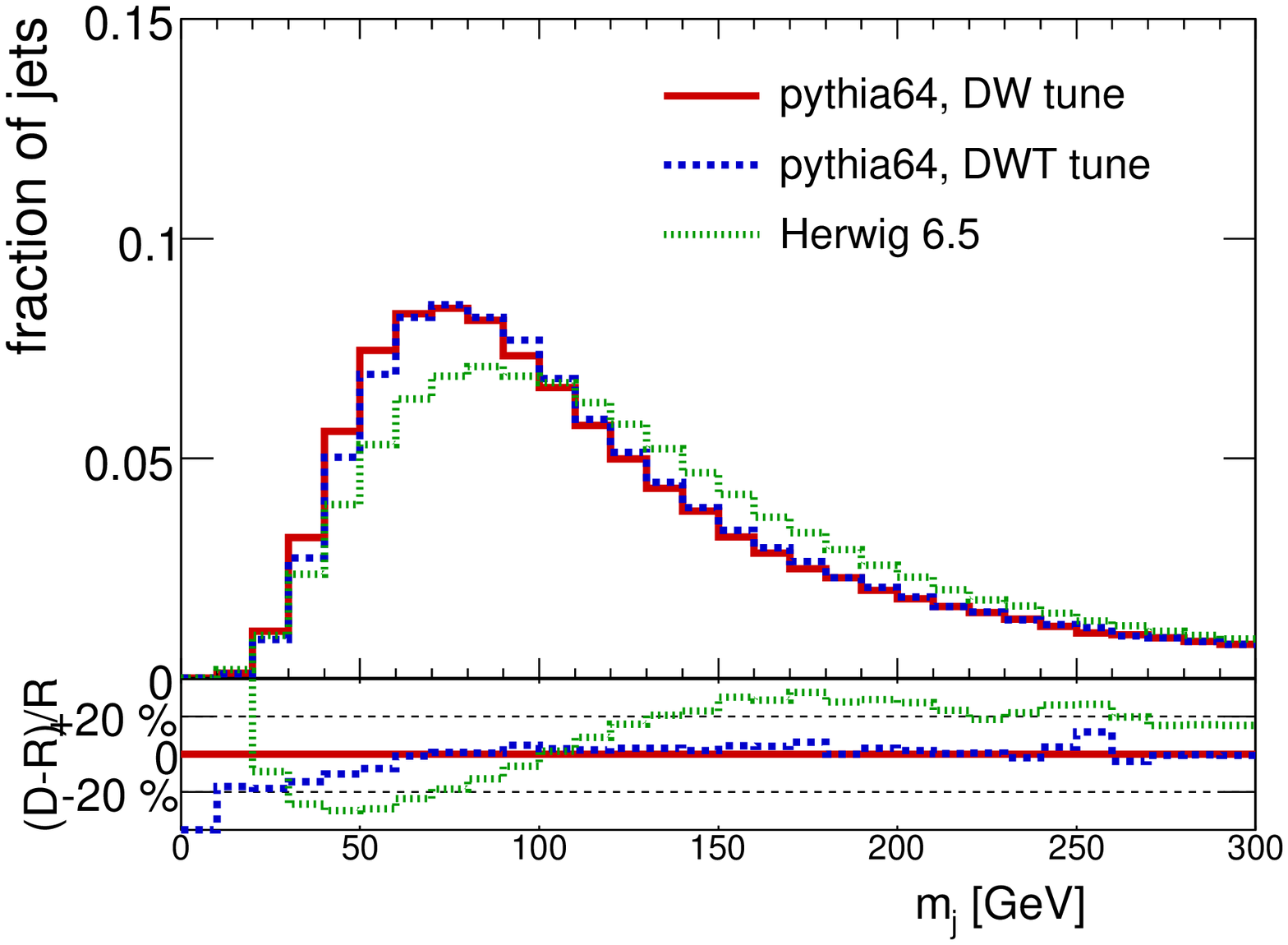}} 
 \subfigure[jet mass - detector]{\includegraphics[width=0.32\textwidth]{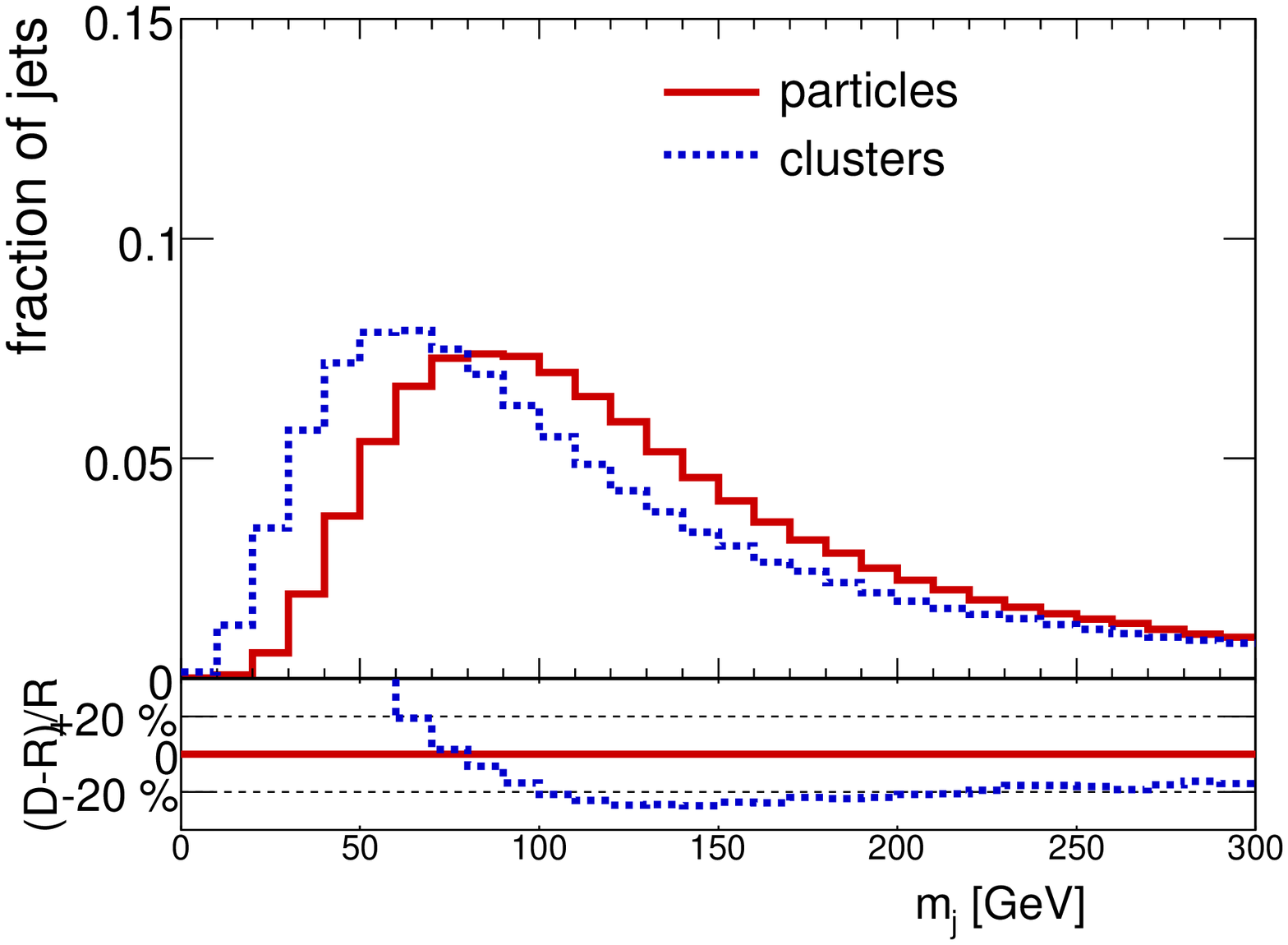}} \\
  \subfigure[groomed jets - PS]{\includegraphics[width=0.32\textwidth]{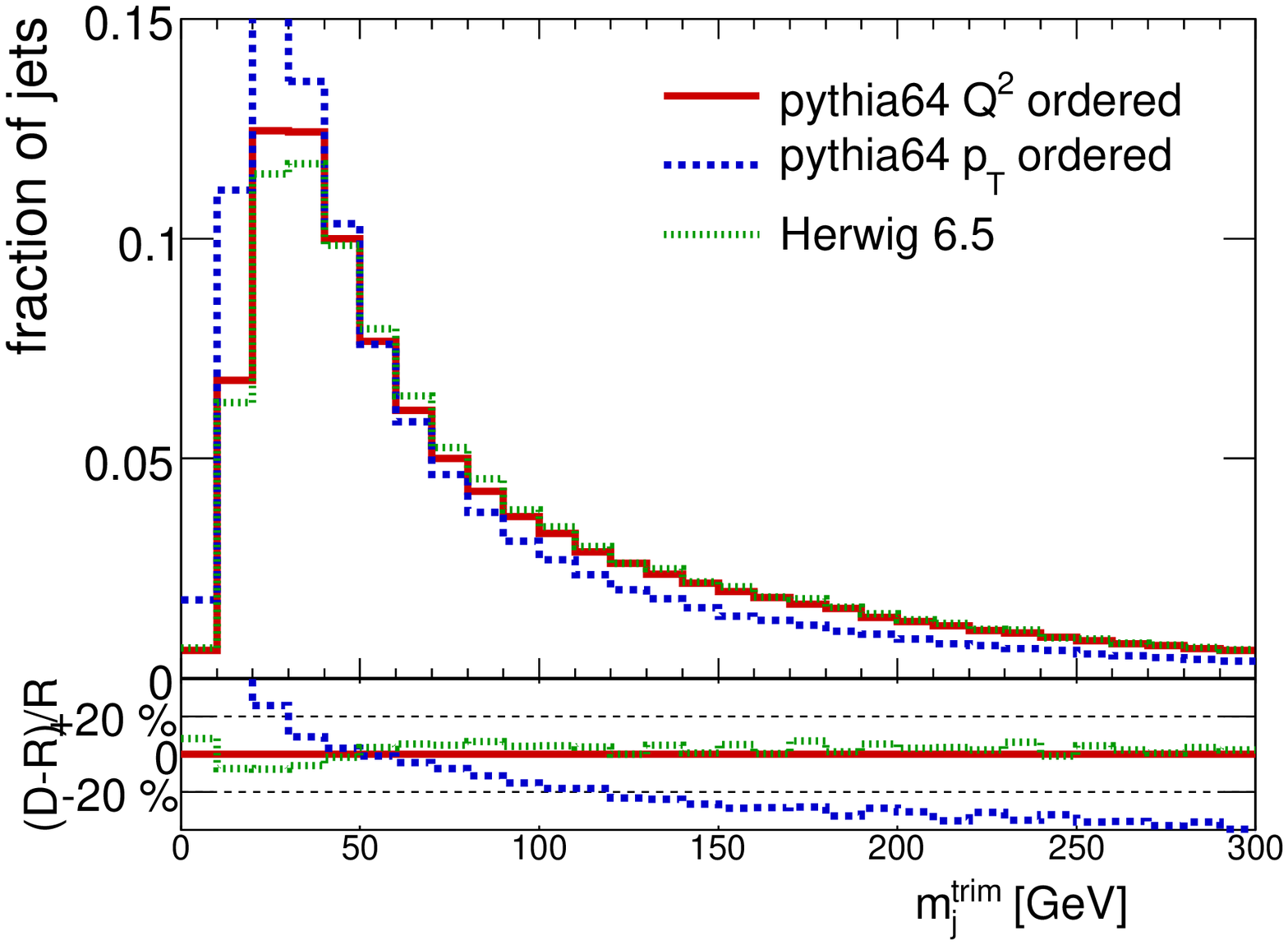}} 
  \subfigure[groomed jets - UE]{\includegraphics[width=0.32\textwidth]{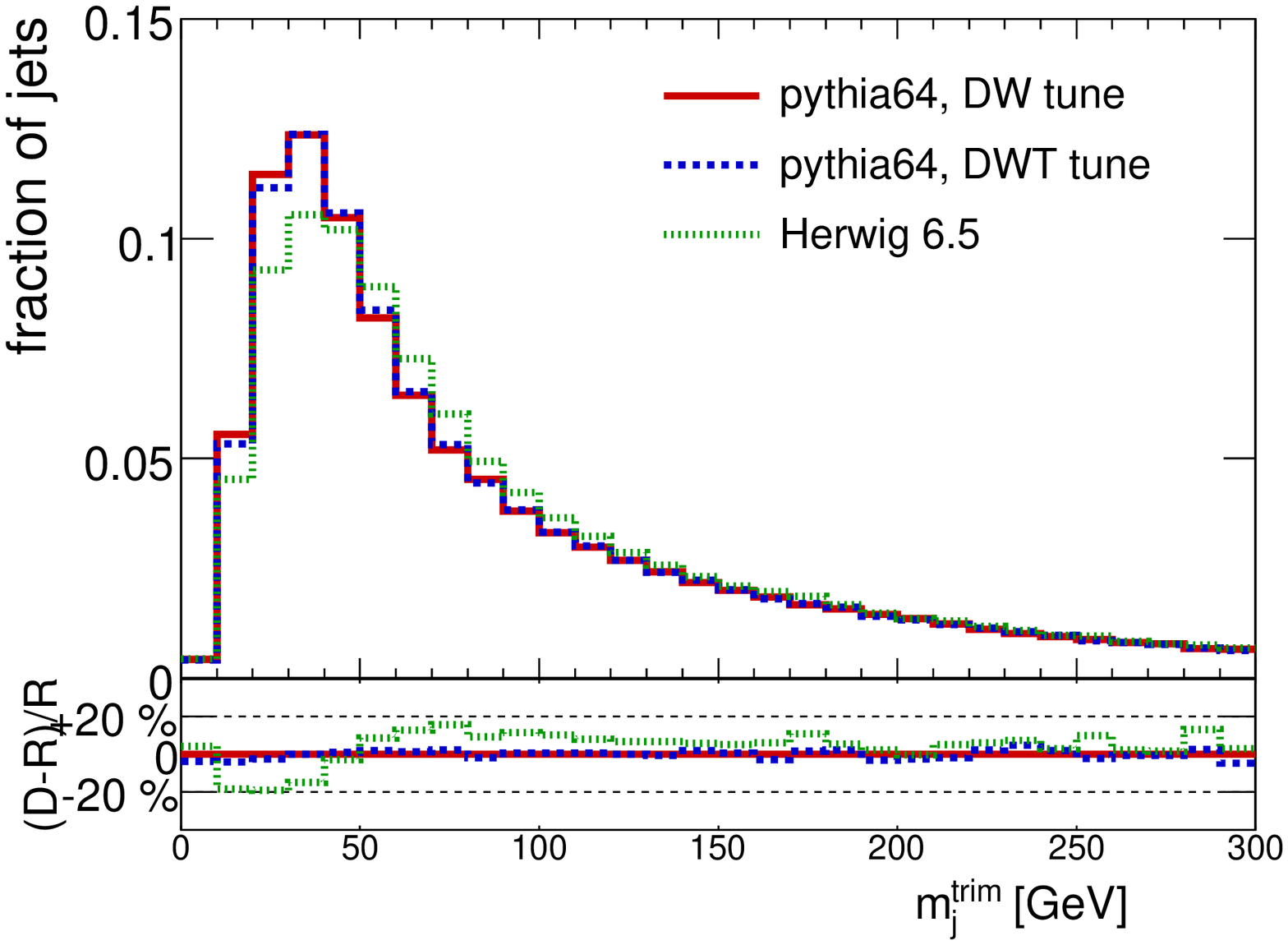}} 
  \subfigure[groomed jets - detector]{\includegraphics[width=0.32\textwidth]{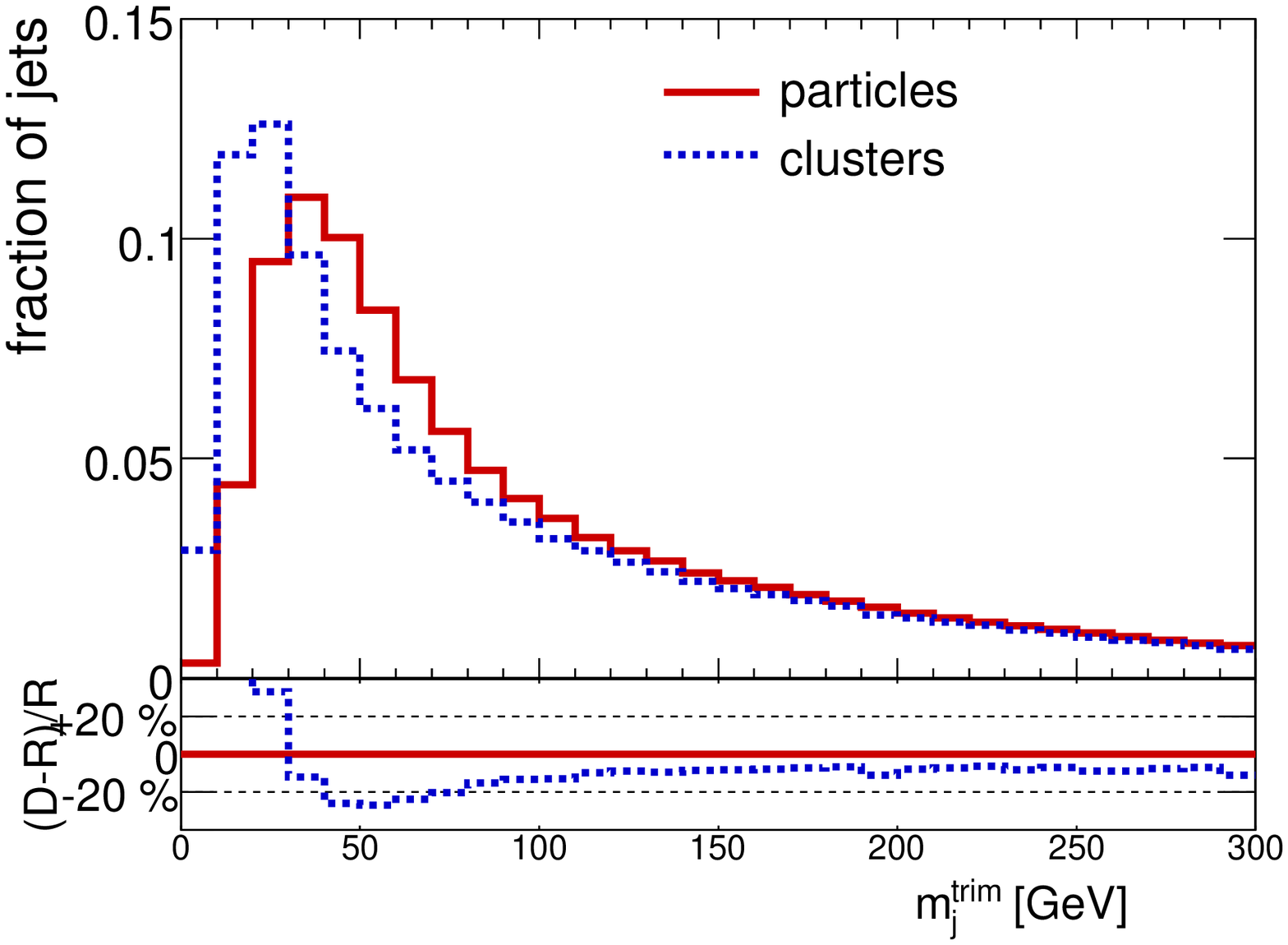}} \\
  \subfigure[$ 1 \rightarrow 2$ split - PS]{\includegraphics[width=0.32\textwidth]{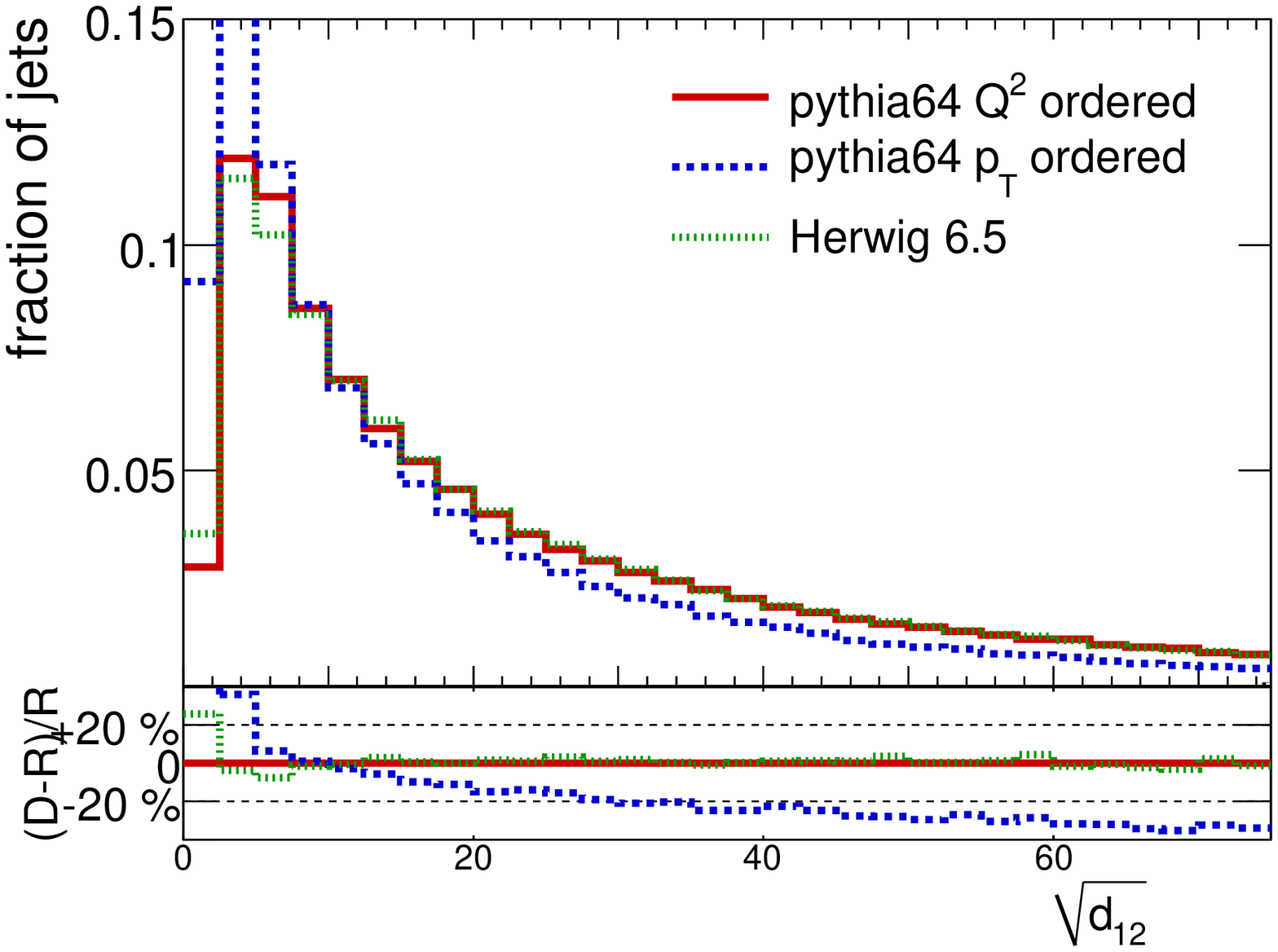}} 
 \subfigure[$ 1 \rightarrow 2$ split - UE]{\includegraphics[width=0.32\textwidth]{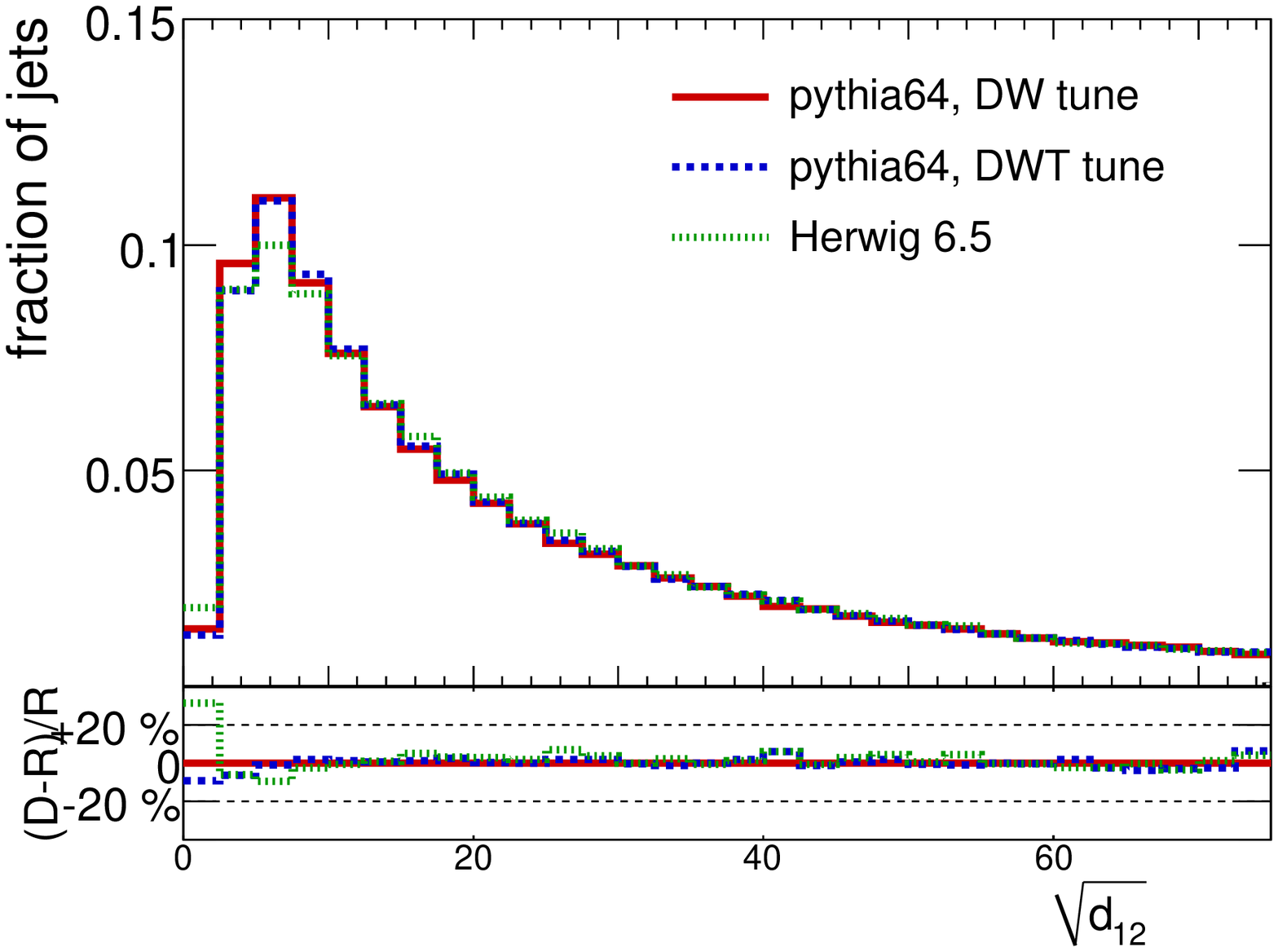}}  
 \subfigure[$ 1 \rightarrow 2$ split - detector]{\includegraphics[width=0.32\textwidth]{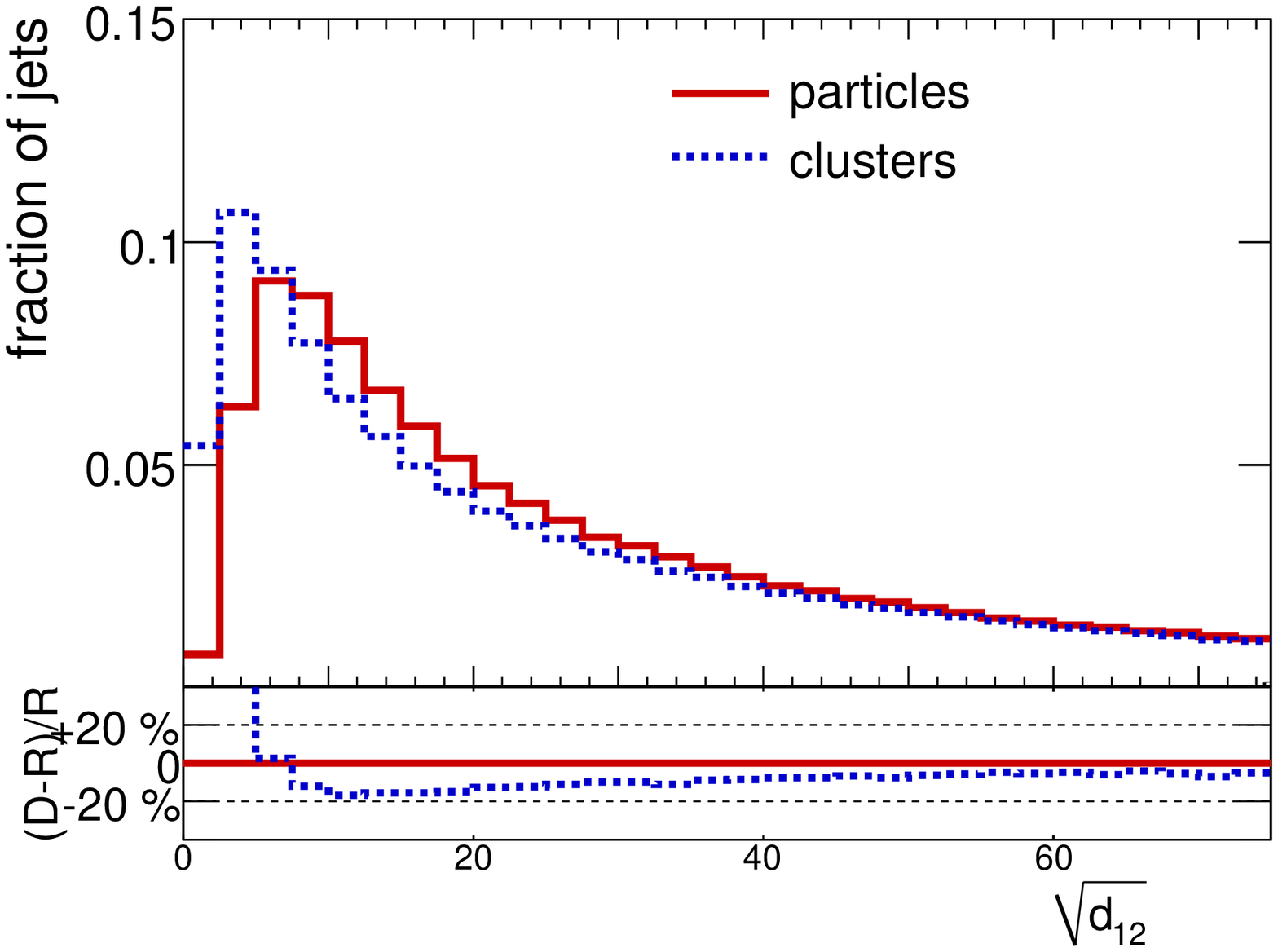}}  \\
  \subfigure[$ 2 \rightarrow 3$ split - PS]{\includegraphics[width=0.32\textwidth]{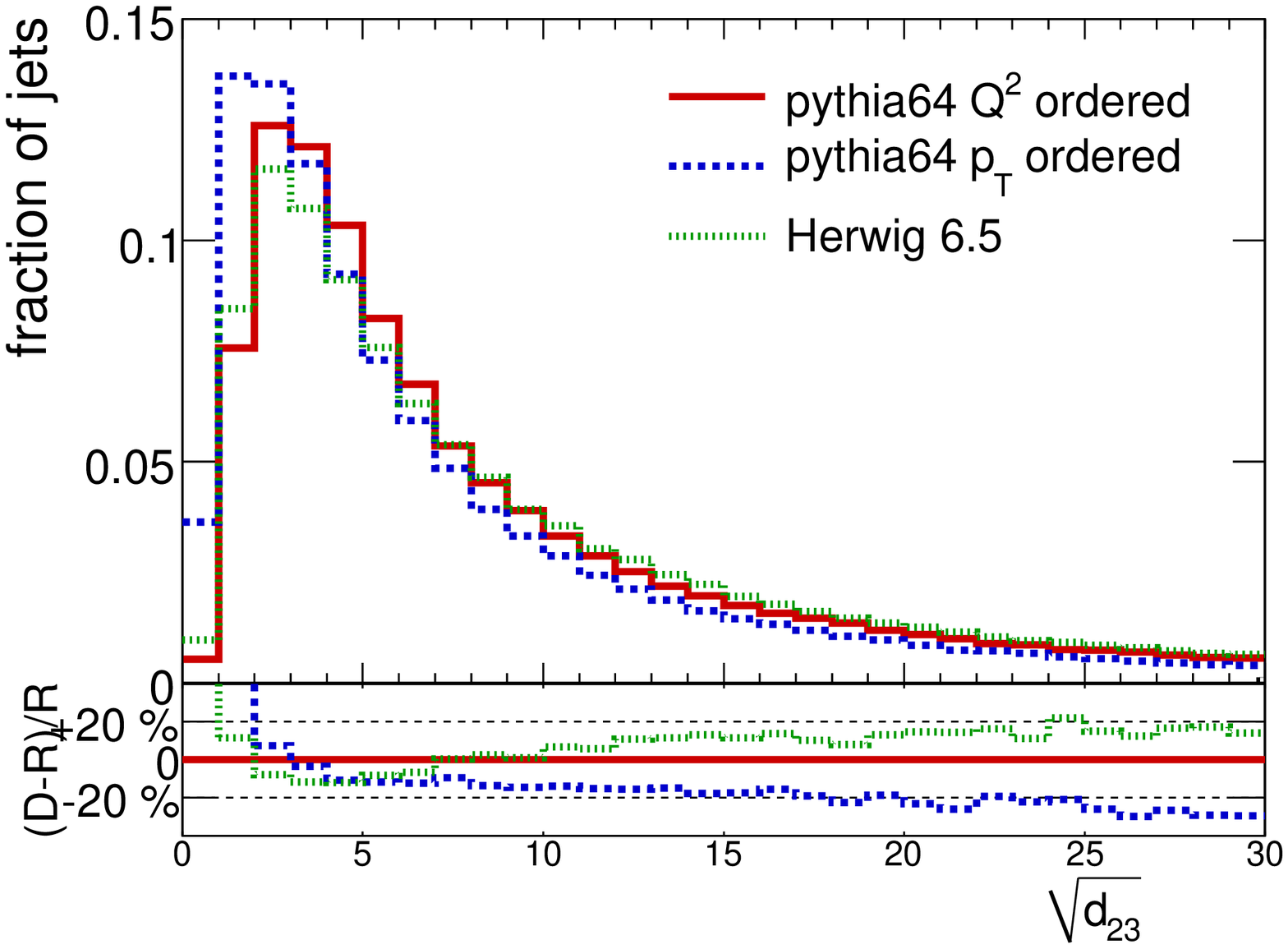}} 
 \subfigure[$ 2 \rightarrow 3$ split - UE]{\includegraphics[width=0.32\textwidth]{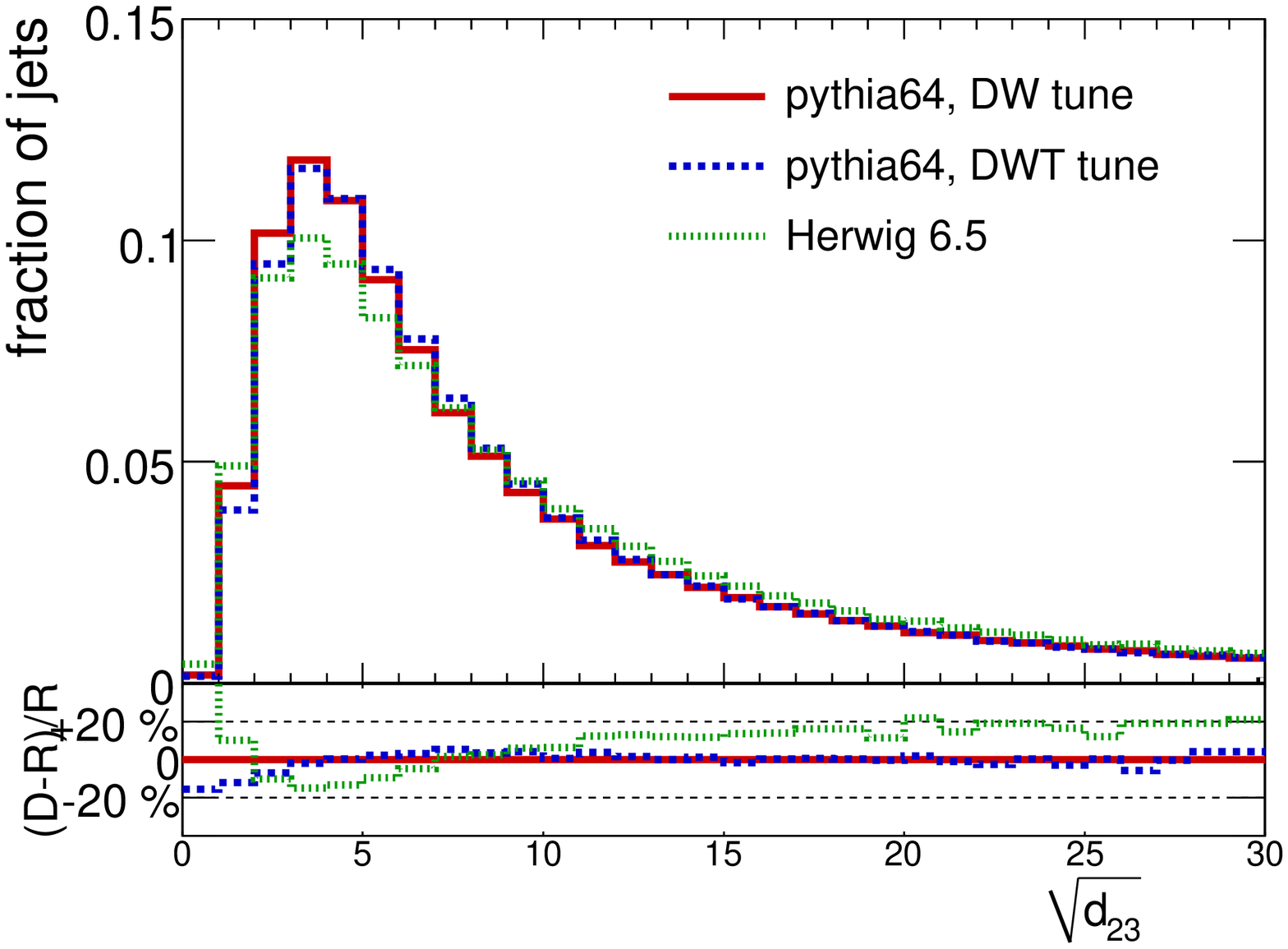}}  
 \subfigure[$ 2 \rightarrow 3$ split - detector]{\includegraphics[width=0.32\textwidth]{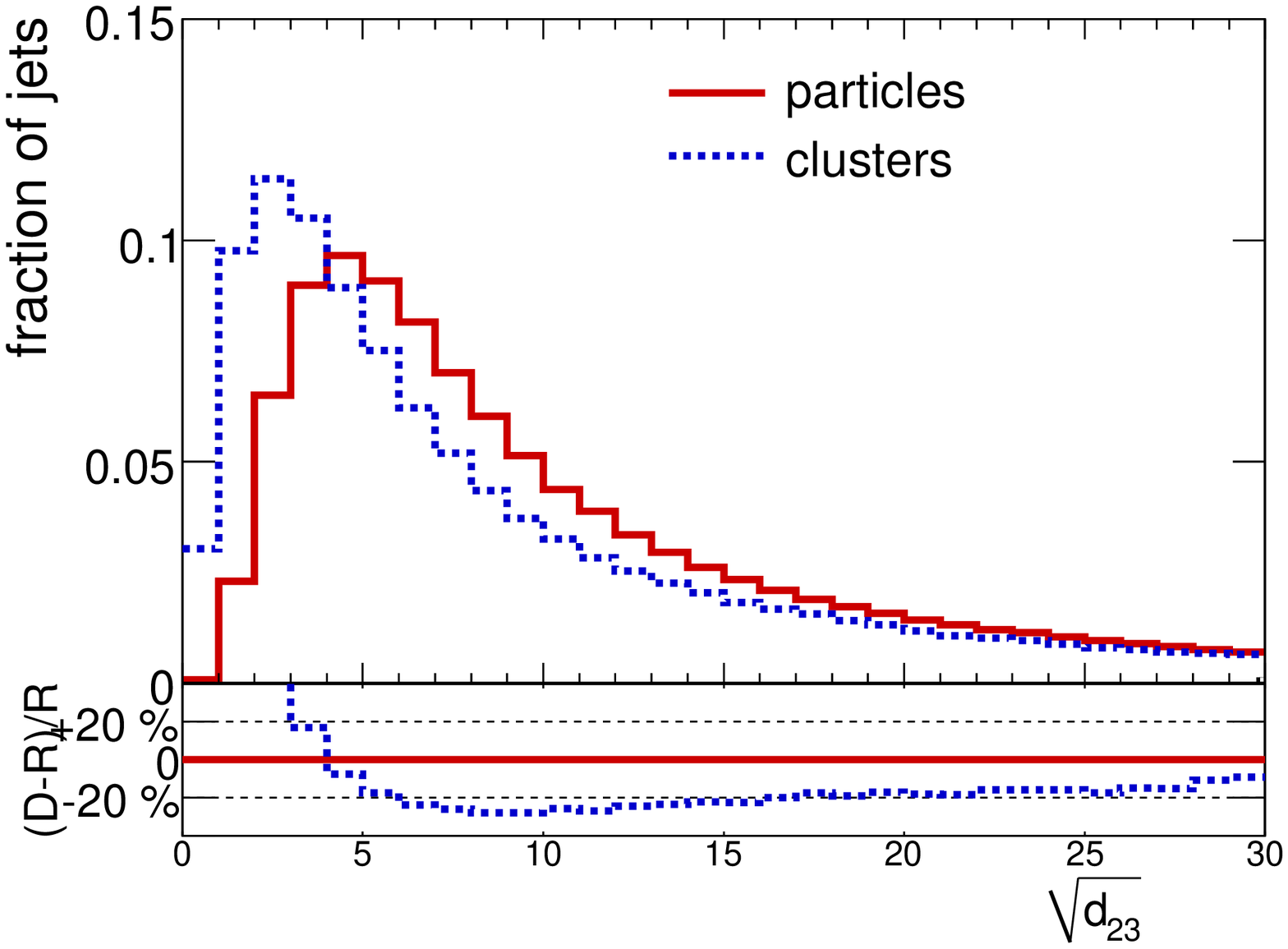}} 
  \end{tabular}
}
\end{center}
\caption{Jet invariant mass $m_j$ before (a-c) and after grooming (d-f), and (ungroomed) splitting scales $\sqrt{d_{12}}$ (g-i) and $\sqrt{d_{23}}$ (j-l) for anti-$k_T$ jets with R=1 reconstructed on dijet samples with an approximately flat distribution in jet $p_T$. The three histograms in the plots of the leftmost column correspond to three different shower models: $Q^2$ and $p_T$ ordered showers in {\sc PYTHIA} with the DW and Perugia tune, respectively, and the default {\sc HERWIG} shower model. In the central column, two {\sc Pythia} underlying event tunes are compared to default {\sc Herwig/Jimmy}. In the rightmost column particle-level jets are compared to cluster-level jets. In the small inset underneath each histogram, the relative deviations from a reference histogram are given ((data-ref)/ref), where the result for {\sc PYTHIA} $ Q^2$-ordered showers (leftmost column), the {\sc PYTHIA} DW tune (central column) and particle level jets (rightmost column) as the reference.} 
\label{fig:had:robust:all}
\end{figure*}

We reconstruct the jet invariant mass distribution for anti-$k_T$ jets with $R=1$. The grooming techniques described in section~\ref{sec:had:comp_groom} select relatively hard events and are therefore expected to reduce the sensitivity to soft and diffuse energy deposits. We apply the three grooming procedures and determine the invariant mass of the resulting groomed jet. We present the result of trimming, but the conclusions hold for all three techniques. We moreover recluster the jet constituents with the $k_T$ algorithm and unwind the sequence to retrieve the $ i \rightarrow j $ splitting scales $d_{ij}$. We note that the splitting scales are determined on the ungroomed cluster sequence.

To establish the impact of different parton shower models we compare the response to two of the most popular Monte Carlo tools for jet formation, {\sc HERWIG} and {\sc PYTHIA}. We moreover vary the order of the emissions in {\sc PYTHIA}, using two schemes known as $p_T$-ordering (used in the Perugia0 tune) and $Q^2$ ordering (used in DW and DWT). In Fig.~\ref{fig:had:robust:all}, we compare the jet mass distribution for these three setups, along with the $k_T$ scales corresponding to the $ 1 \rightarrow 2 $ and $ 2 \rightarrow 3 $ splits. For the sake of a clean comparison we disabled UE activity for these samples.

We find the $p_T$ ordered shower in {\sc PYTHIA} yields a significantly softer spectrum than the $ Q^2$ ordered shower model. This is true for the jet invariant mass and the scales of the hardest splittings in the shower. The results obtained for the {\sc HERWIG} shower are in good agreement with the $ Q^2$ ordered shower for both the jet mass and the $ 1 \rightarrow 2 $ splitting scale. 

We expect larger differences between Monte Carlos in the region of larger 
masses and splitting scale, as these probe less collinear regions of the 
jet structure, where the codes are less constrained. Unfortunately matching to fixed-order prediction (such as with Alpgen or Madgraph) may not necessarily be of immediate help in this context given the recent results~\cite{cmsshapes} which show that matching does not necessarily improve the description of the event structure near the dijet limit.

For the $ 2 \rightarrow 3 $ splitting scale we observe relative differences of up to 20 \%. The greater robustness of the $ 1 \rightarrow 2 $ splitting scale compared to the $2 \rightarrow 3$ splitting scale might have been expected. None of the Monte Carlos explicitly includes an exact collinear $ 2 \rightarrow 3$  splitting kernel, whereas they do all include the $ 1 \rightarrow 2$ kernel.

In the experimental environment, jet observables are affected by UE activity and energy flow due to pile-up events. 
These effects are particularly important for the large jet sizes envisaged for many searches. It is therefore important to establish the sensitivity of substructure analyses to such effects. In Fig.~\ref{fig:had:robust:all} we compare the distributions for the same three observables for three different UE tunes.

The larger UE activity in DWT with respect to the DW tune in {\sc PYTHIA} is reflected in a (slightly) increased jet mass. The {\sc HERWIG}+{\sc JIMMY} jet mass spectrum is significantly harder than that of either {\sc PYTHIA} tune. 
Although the UE activity is typically soft, it can have a sizable effect on the invariant mass of the jet. This deviation is clearly observed in all $p_T$ bins from 200~\gev to 1.5~\tev. For the first splitting scale, on the other hand, we find excellent agreement between the three tunes. 
Our interpretation is that this observable corresponds to the hardest event in the shower development and is therefore least sensitive to unrelated, soft activity. 
This is consistent with our observation that consecutive, softer, splittings ($ 2 \rightarrow 3 $ and $ 3 \rightarrow 4 $) exhibit an increasing discrepancy between the {\sc PYTHIA}~ DW and {\sc HERWIG} distributions.

Finally, the measurement of substructure observables will be affected by detector limitations. We study two important effects here by comparing particle and cluster level results and leave the remainder for future studies. In our simple setup, the detector granularity is simulated by forming massless clusters that contain the energy of all particles in a $y - \phi$ region of $ 0.1 \times 0.1 $. A 1~\gev threshold is applied to the resulting cluster $E_T$.

The jet mass, in Fig.~\ref{fig:had:robust:all}, is found to be quite sensitive to detector 
effects. The peak of the distribution for the QCD jet background is
shifted down by several tens of GeV. The same is found to be true for the $t \bar{t}$ signal. 
The groomed mass distribution is much less affected. 
The splitting scales are found sensitive to detector effects, in the region  $\sqrt{d_{ij}} < $ 20~\gev.


To summarise, we have investigated the sensitivity of some of the most popular jet substructure observables to uncertainties in the MC description of the parton shower, the UE and the detector response. We find that observables envisaged to be used in the selection of new physics are strongly affected by some of these effects. The MC samples for evaluating the performance of new algorithms must therefore be chosen carefully. We recommend the benchmark samples presented in section~\ref{sec:had:samples} be used to provide a comparison under equal conditions.

The significant difference we observe between different MC tools suggest there 
is benefit to be had by more extensive measurements of a range of jet subtructure and shape observables at the LHC, where the very high statistics available at moderate $p_t$ could provide strong additional constraints on the generators. 
For early comparisons of different shower models and tunes to LHC data the reader is referred to recent studies of jet shapes by ATLAS~\cite{atlasjetshapes} and CMS~\cite{cmsjetshapes}.

We find that the investigated observables show rather differing sensitivities. 
We find the invariant mass of the jet to be quite sensitive to UE activity and 
detector effects. The grooming techniques investigated in this paper greatly 
improve the robustness of the jet mass. Also the $k_T$ splitting scales are 
quite robust, provided their use is limited to the region above approximately 
20~\gev. For the two most commonly used shower models, all observables are in 
good agreement, but the $p_T$ ordered shower in {\sc PYTHIA} yields 
significantly different results. 



\section{Comparison of top-tagging tools}
\label{sec:had:comp}



\begin{figure*}[t!]
\begin{center}
\mbox{
  \begin{tabular}{cc}
  \subfigure[all \pt samples] {\includegraphics[width=0.48\textwidth]{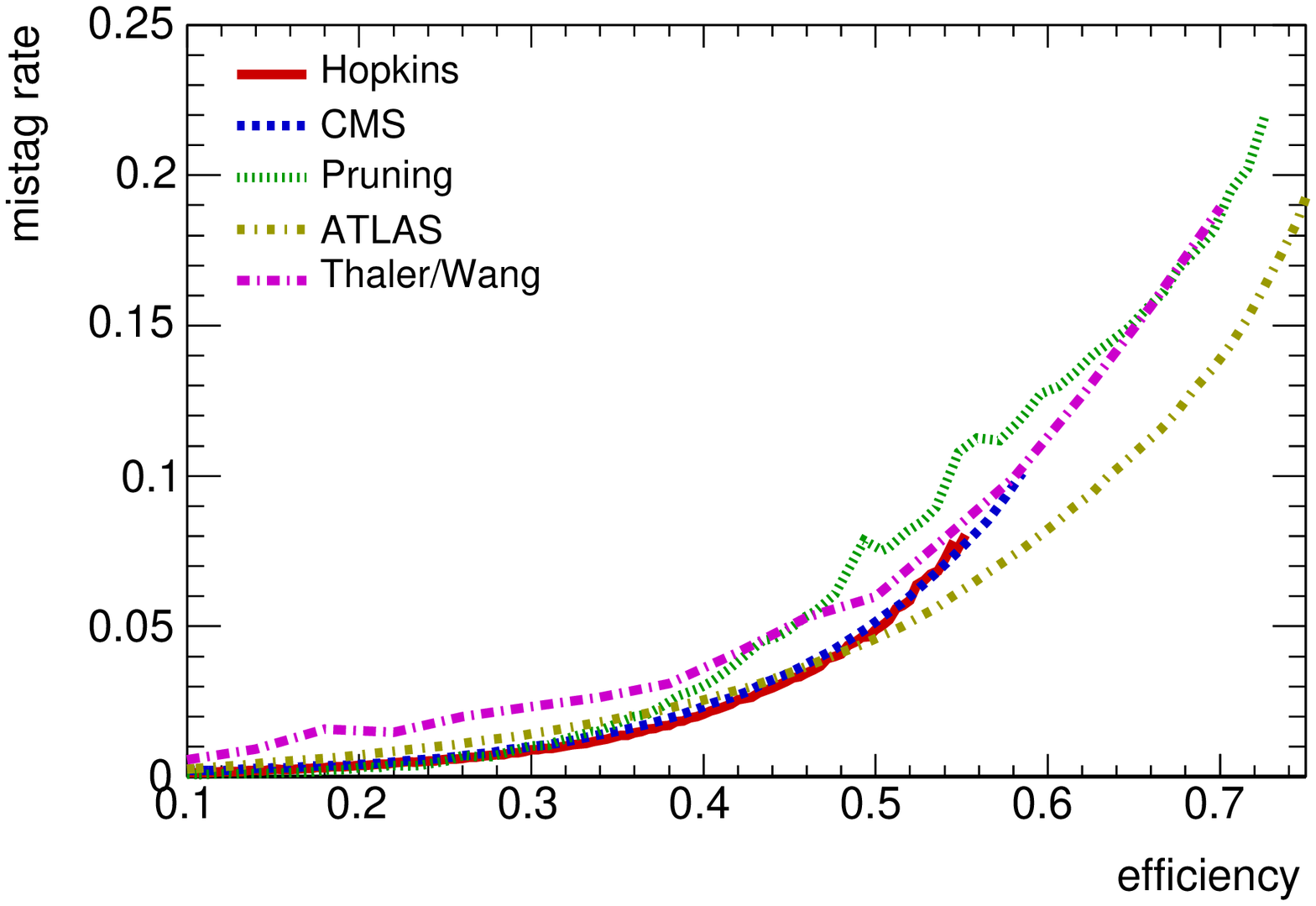}}
  \subfigure[all \pt samples] { \includegraphics[width=0.48\textwidth]{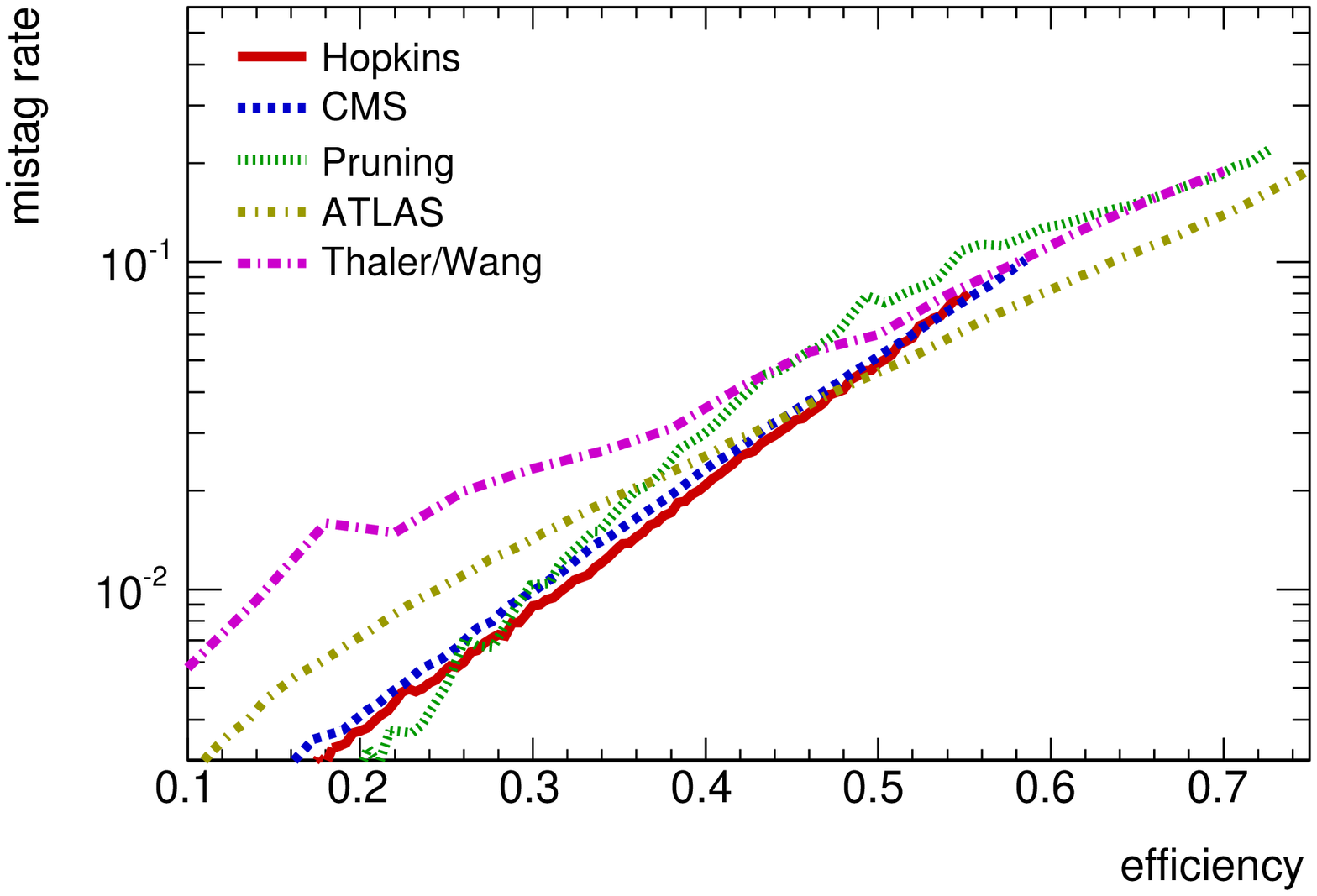}} \\
  \subfigure[300--400 \GeV] { \includegraphics[width=0.48\textwidth]{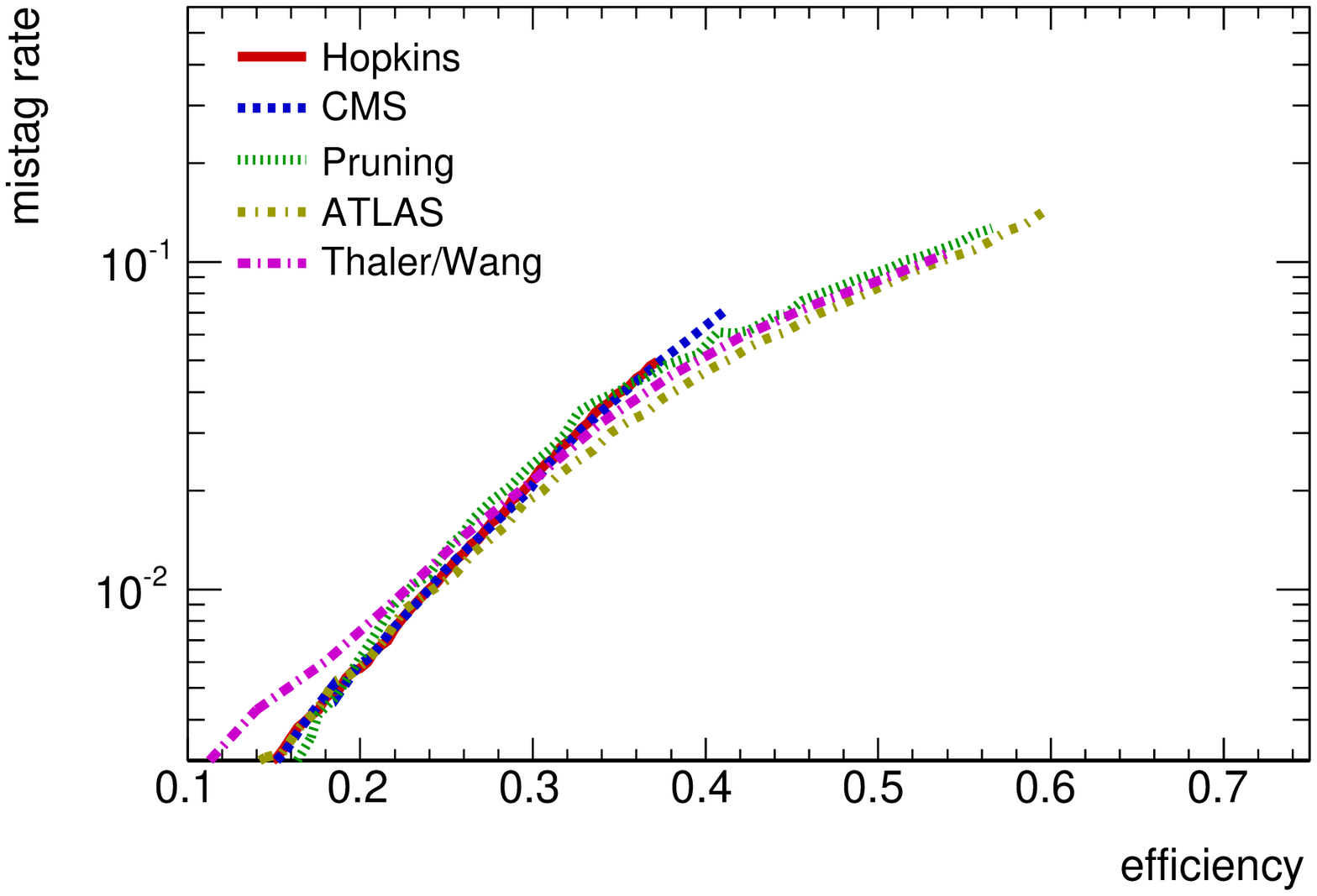}}
  \subfigure[500--600 \GeV] { \includegraphics[width=0.48\textwidth]{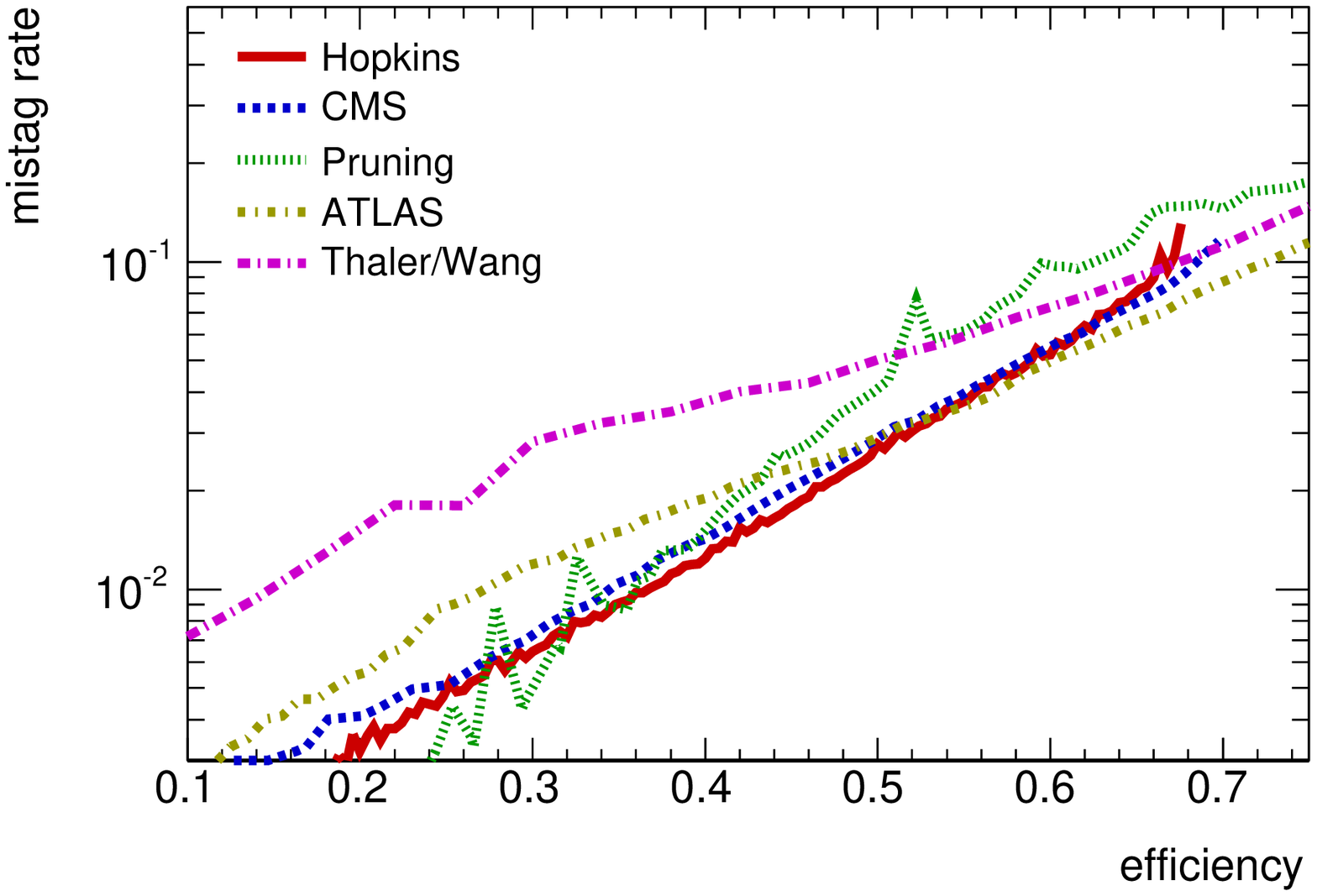}}
 \end{tabular}
}
\end{center}
  \caption{Mistag rate versus efficiency after optimisation
  for the studied top-taggers in linear scale (a) and logarithmic scale (b).
  Tag rates were computed averaging over all \pt subsamples (a,b) and for the subsample containing jet with \pt range 300--400 \GeV (c) and 500--600 \GeV (d)}
\label{fig:hadroniccomp:mistag_eff}
\end{figure*}

We have performed a study to compare the different top-tagging algorithms.\footnote{Several interesting approaches to top-tagging are not included in this study. The complexity of the template-based top tagger of Reference~\cite{Almeida:2010pa} precluded its inclusion on the timescale of this study. Another promising new approach, known as subjettiness~\cite{Thaler:2010tr} was published only after the workshop. We hope the performance measurement of these new approaches can be included in future work.}
 
The benchmark QCD dijet (background) and  \ttbar (signal) samples, produced using {\sc HERWIG} as described in section \ref{sec:had:samples}, were used. 
However, for this study, only the subsamples with
parton $\pt$ ranges up to and including 700--800 \GeV bin were used. 
For each event, jets were clustered with the anti-$k_T$ algorithm with an $R$-parameter of 1.0. 

For each anti-$k_T$ jet, top-tagging algorithms are run on the constituents of the jet. As the final step of top-tagging,
all algorithms applied selection criteria on kinematic variables such as jet mass and jet substructure. Applying the
top-tagging algorithms with their default cut values yields different mistag rates and efficiencies which
makes it difficult to compare them directly. In our study, for each top-tagging algorithm, the cuts were optimised for each efficiency by minimising the mistag rate while keeping the efficiency fixed.
In this context, the overall mistag rate and overall efficiency are defined as the number of top-tags divided by
the total number of anti-$k_T$ jets in the background and signal sample, respectively. The normalisation uses anti-$k_T$ jets above 200~\GeV and at most two per event.

As can be seen from Fig.~\ref{fig:had:comp_groom:mjet} in Section~\ref{sec:had:comp_groom}, jets at low $p_T$ values often have an invariant mass inconsistent with the top quark mass. These jets
include only some decay products of the hadronic top quark decay, e.g., the quarks of
the hadronic $W$ boson decay. Running the optimisation procedure in this $p_T$ region would hardly result
in a top-tagger but possibly rather a ``$W$-tagger''. Therefore, we impose an additional cut on the
anti-$k_T$ jet mass of $m_{\mathrm{jet}} > 120\GeV$ for all top-taggers. This implies
a maximum overall tagging efficiency of 75\%.

Curves with the optimal mistag rate versus signal efficiency are shown in
Fig.~\ref{fig:hadroniccomp:mistag_eff}.
The optimisation was repeated on the \pt subsamples and can be compared to the overall optimisation applied on the subsample
to evaluate the potential benefit of using \pt-dependent cut values. Curves for the $300 < p_T < 400\GeV$ (c) and $500 < p_T < 600\GeV$ (d) subsamples are also shown.

While these curves can be used to compare the overall performance of the top-tagging algorithms, they do not reflect the \pt-dependence of the tag rate. We expect that, at least initially, the experiments are likely to choose a
  single set of parameters across the whole \pt range in order to keep
  their analyses of these new tools as simple as possible. It is therefore 
instructive to look at the tag rate as function of jet \pt for specific 
working points. We chose two working points defined by their overall signal 
efficiency of 20\%  and 50\%.


Firstly, we investigated the performance of two taggers that do not incorporate 
any grooming procedures. The first one is referred to as the ATLAS 
tagger~\cite{ATL-PHYS-PUB-2010-008,brooijmans2,brooijmans}, the second one as 
the Thaler/Wang (T/W) tagger~\cite{Thaler:2008ju}. Both of them exploit the inherent 
hierarchical nature of the $k_T$ jet algorithm by reclustering the initial 
jet's constituents. The final and penultimate 
stages of this process correspond on average to the merging of the top quark 
decay products and hence jet substructure can be probed via the first few $k_T$
 splitting scales.

\begin{figure*}[t!]
\begin{center}
\mbox{
  \begin{tabular}{cc}
 \subfigure[ATLAS, Thaler/Wang, $\epsilon = $ 20 \%]{ \includegraphics[width=0.32\textwidth]{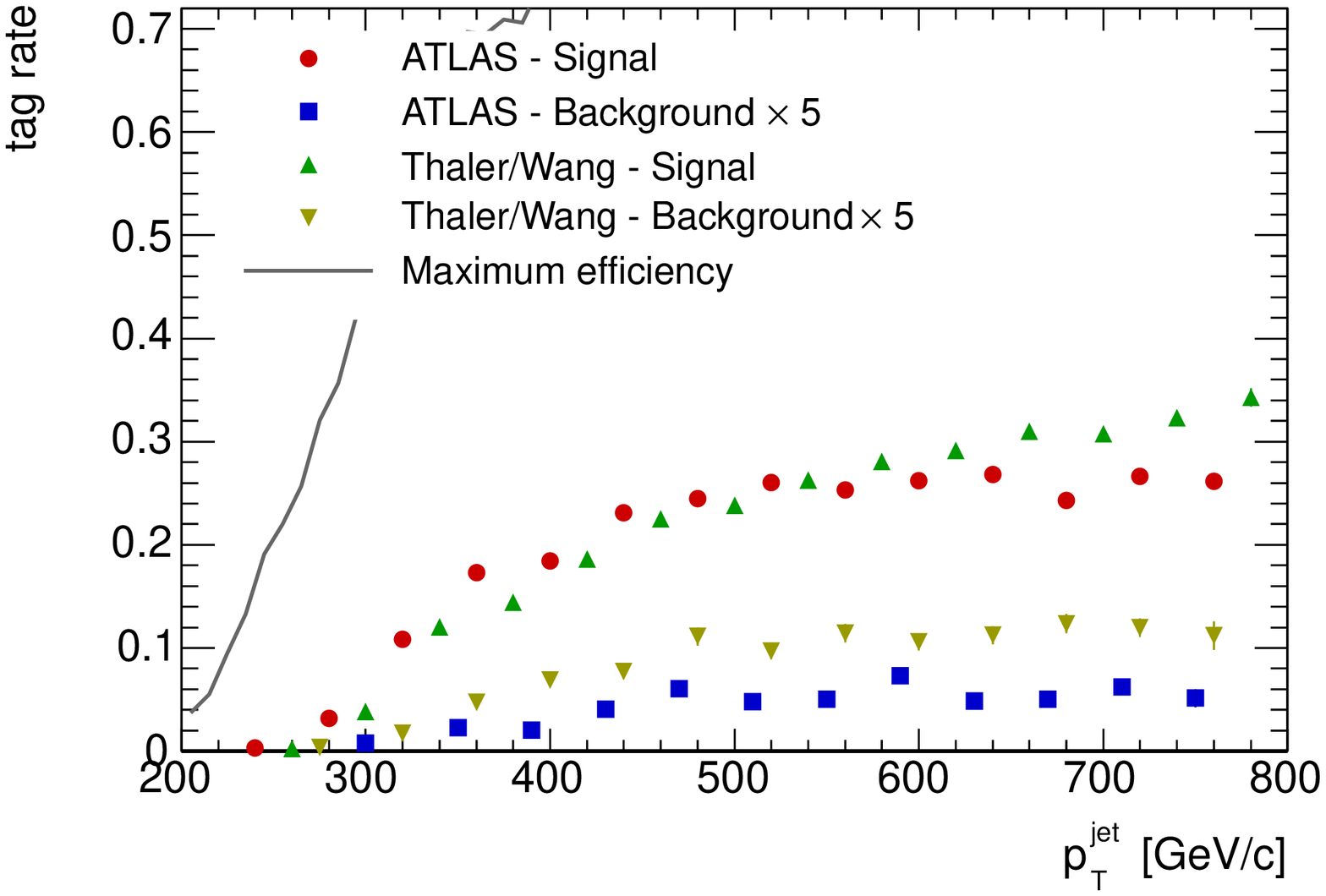}} 
\subfigure[CMS, Hopkins, $\epsilon = $ 20 \%]{  \includegraphics[width=0.32\textwidth]{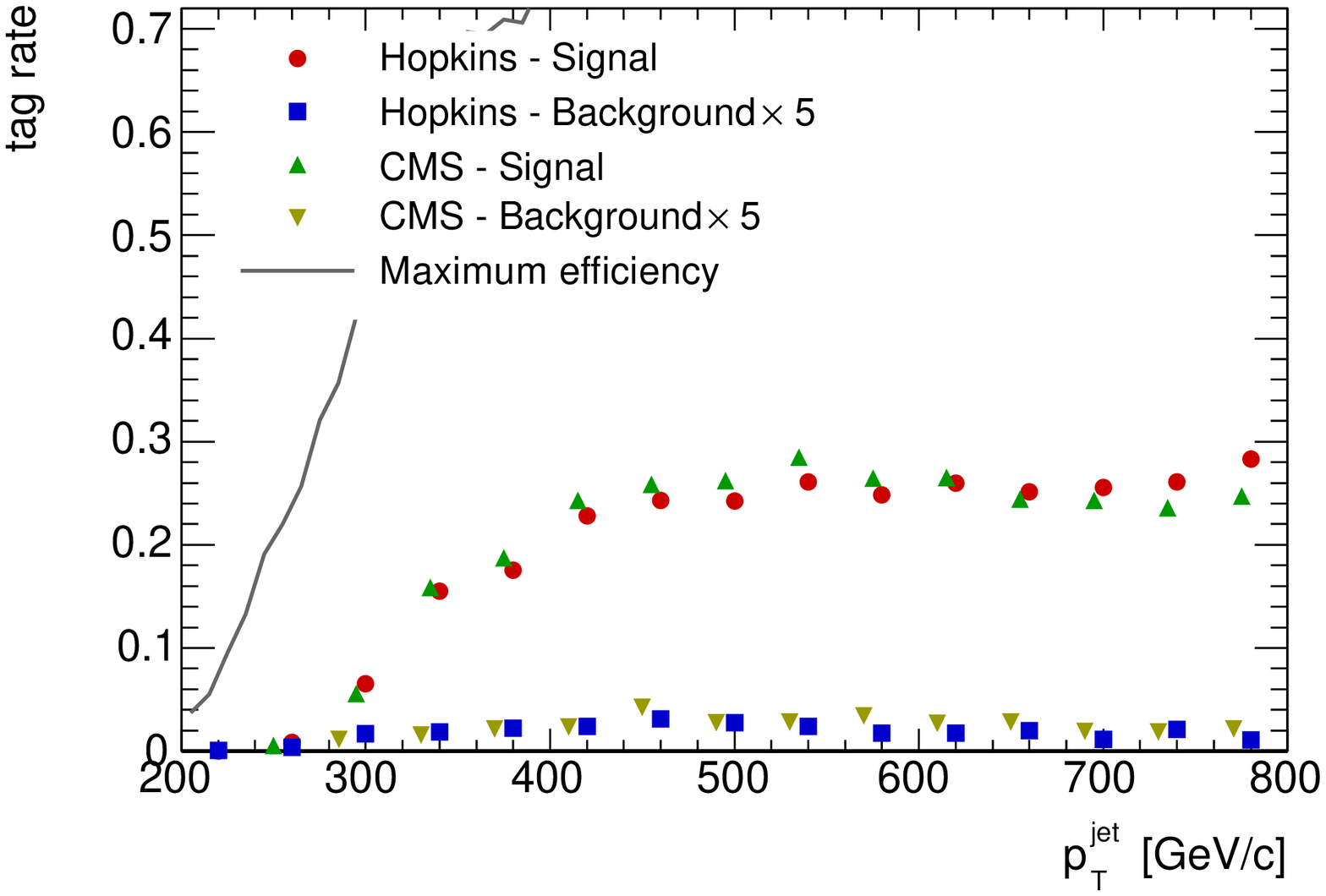}} 
\subfigure[pruning, $\epsilon = $ 20 \%]{ \includegraphics[width=0.32\textwidth]{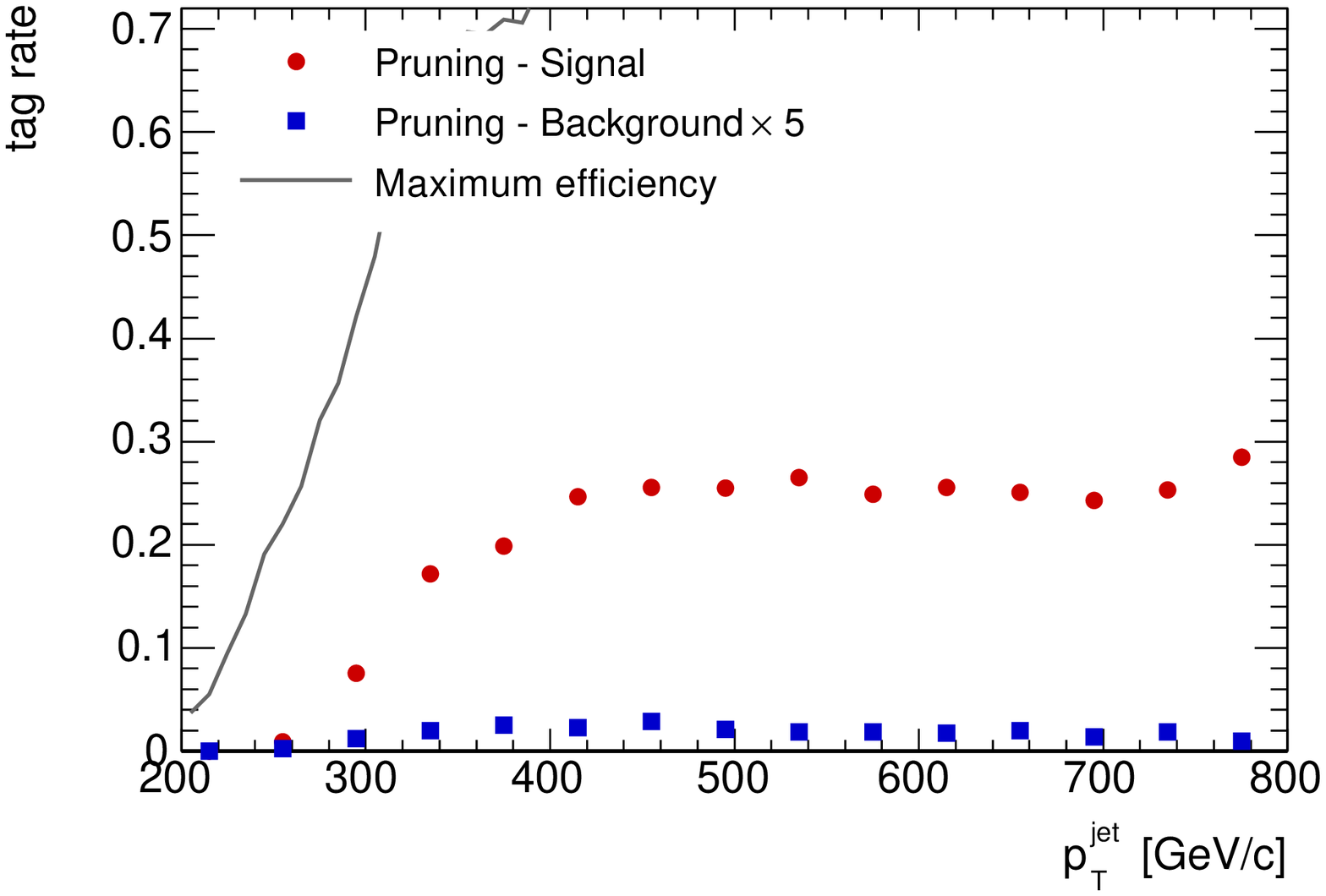}} \\
  \subfigure[ATLAS, Thaler/Wang, $\epsilon = $ 50 \%]{ \includegraphics[width=0.32\textwidth]{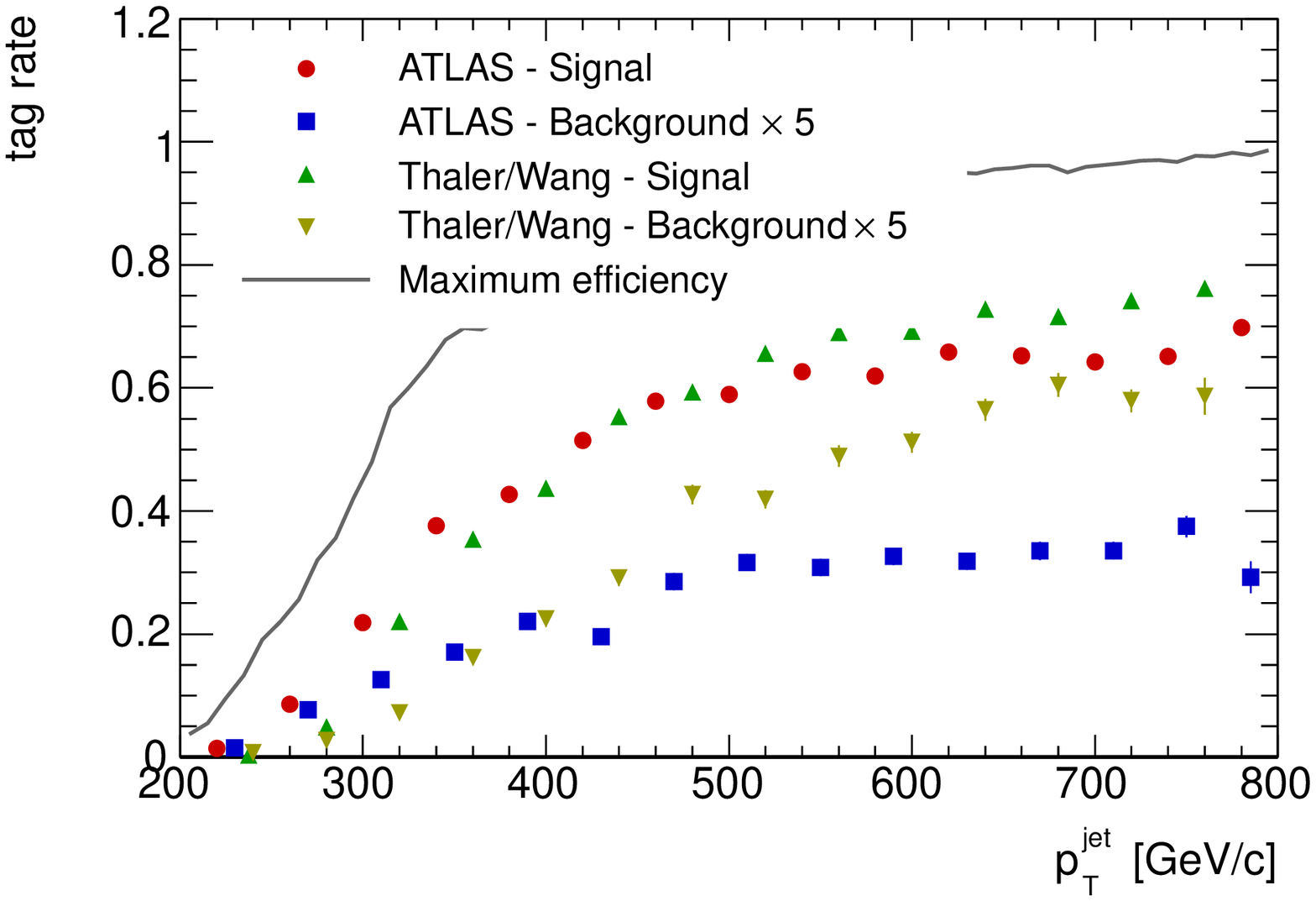}} 
  \subfigure[CMS, Hopkins, $\epsilon = $ 50 \%]{  \includegraphics[width=0.32\textwidth]{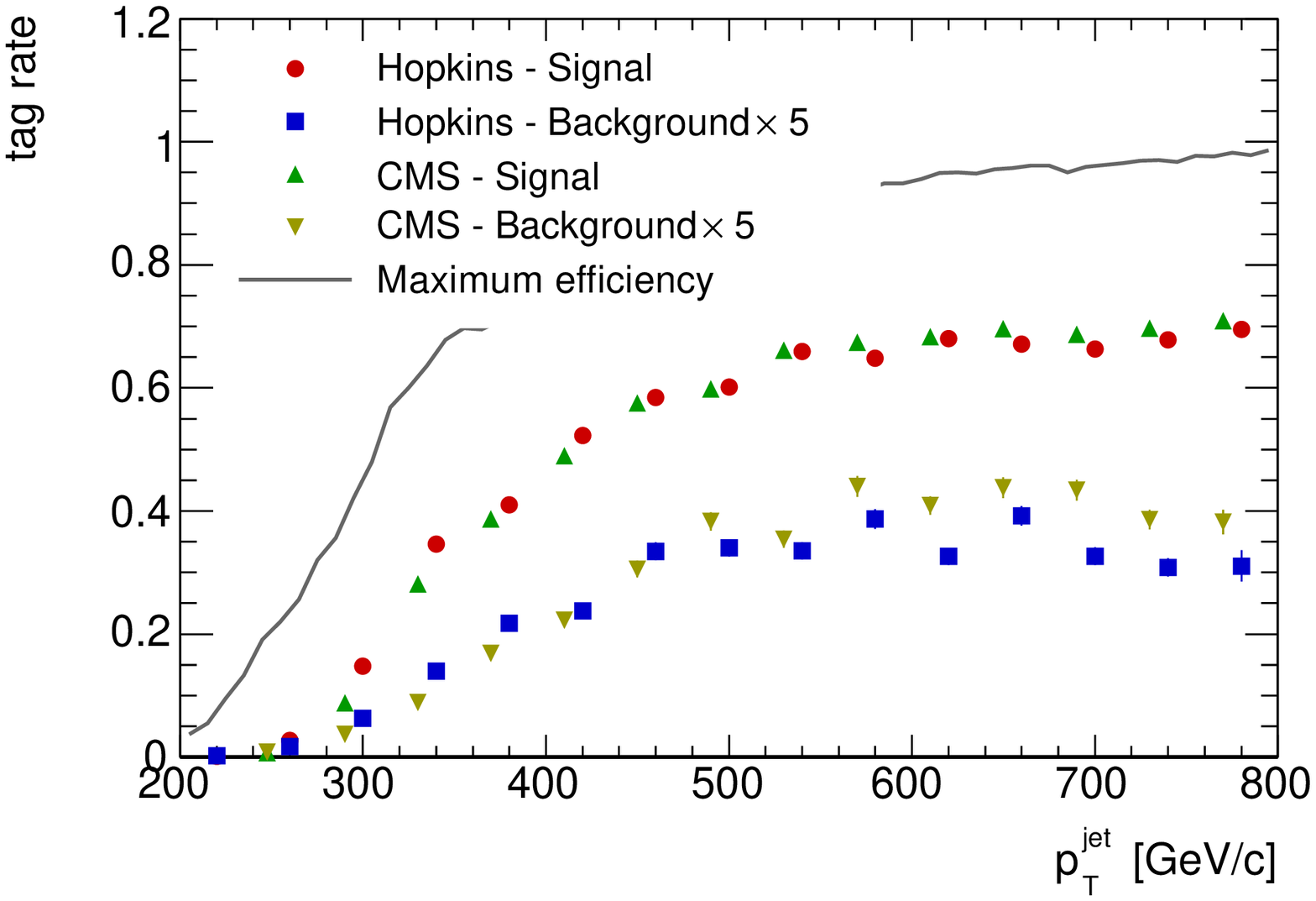}} 
 \subfigure[pruning, $\epsilon = $ 50 \%]{  \includegraphics[width=0.32\textwidth]{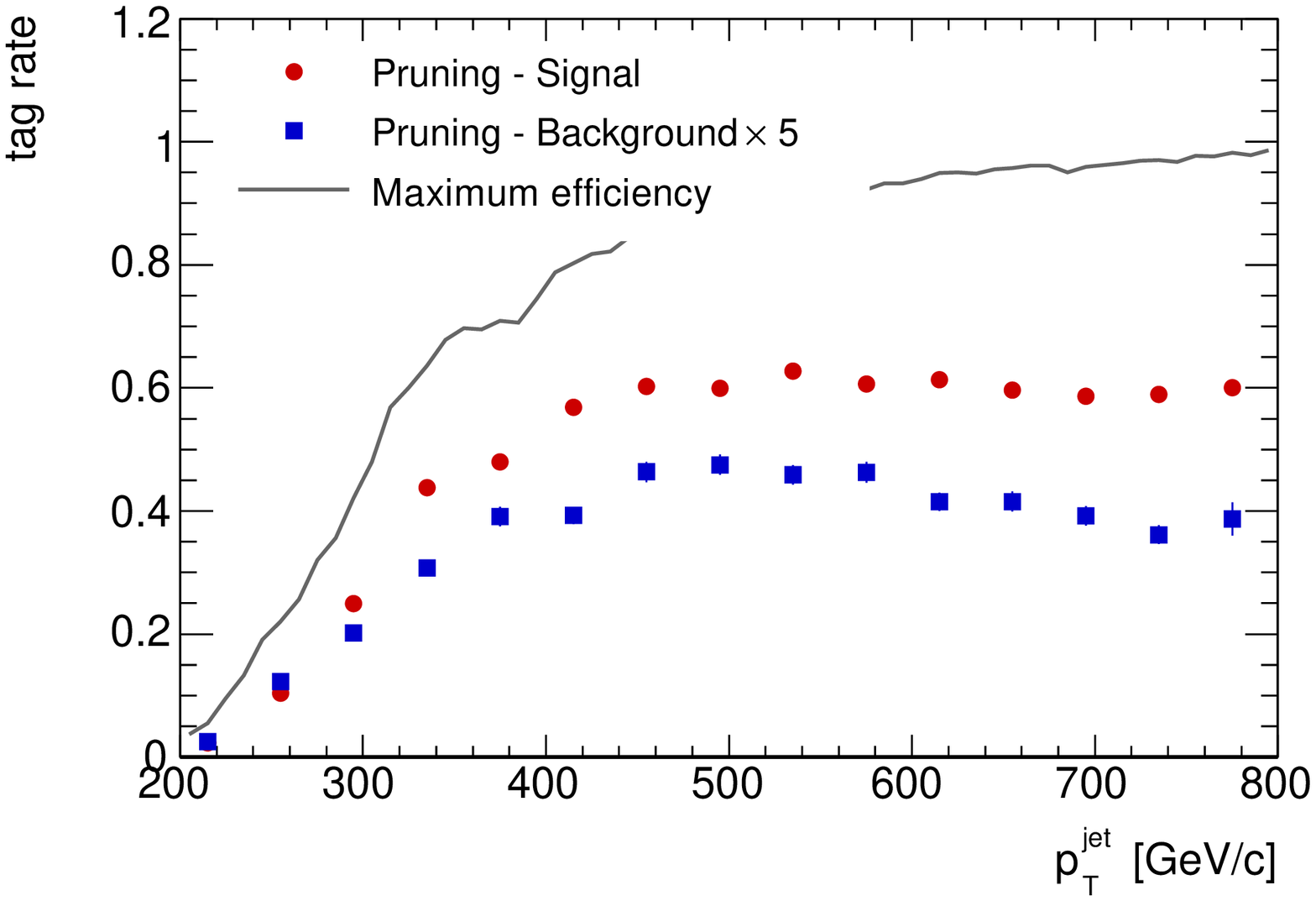}}
 \end{tabular}
} 
\end{center}
\caption{Efficiency and mistag rate as function of jet \pt
  for working points with overall efficiency of 20\% (uppermost row) and 50\% (lowermost row). Results correspond to the ATLAS and Thaler/Wang taggers (a,d), the Hopkins and CMS taggers (b,e) and the pruning tagger (c,f). The mistag rate has been multiplied by a factor 5 to make it visible on the same scale.}
\label{fig:hadroniccomp:tagrate_pt}
 \end{figure*}

\begin{table*}[htb!]
\begin{center}
\begin{threeparttable}
\begin{tabular}{|c|c|c|}
\hline
Tagger   & Parameters at 20\% working point &  Parameters at 50\% working point \\
\hline
\hline 
      & $\delta_p =0.1$, $\delta_{r}=0.19$    &  $\delta_p =0.04$, $\delta_{r}=0.19$ \\
Hopkins      & $ 170  < m_{\text{top}} < 195 \GeV$,    & $ 160  < m_{\text{top}} < 265 \GeV$,   \\
      & $\cos{\theta_h} < 0.675$ , $ 75 < m_W < 95 \GeV$    &   $\cos{\theta_h} < 0.95$, $ 60 < m_W < 120 \GeV$  \\
\hline
          & $170 < m_{\mathrm{jet}} < 200\GeV$    & $164 < m_{\mathrm{jet}} < 299\GeV$   \\
\raisebox{0.5ex}{CMS}          & $ m_{\mathrm{min}} > 75\GeV$    &  $ m_{\mathrm{min}} > 42.5\GeV$  \\
\hline
&  $z_\text{cut}=  0.1$,  $D_\text{cut}/(2m/p_T)= 0.2$  & $z_\text{cut}=  0.05$, $ D_\text{cut}/(2m/p_T)= 0.1$   \\
Pruning\tnote{a}    &   $68 <m_W < 88 \gev $   &   $28 <m_W < 128 \gev $     \\ 
    &   $ 150 <m_\text{top}< 190 \gev$  &  $ 120 <m_\text{top}< 228 \gev$   \\
\hline
ATLAS\tnote{b} &  N/A  &  N/A  \\
\hline
        & $m_W> 68 \gev$  &  $m_W> 59 \gev$  \\
Thaler/Wang        &  $ 0.249 < z_{\mathrm{cell}}< 0.664 $  &  $ 0.0498 < z_{\mathrm{cell}}< 0.509 $  \\
        & $ 183  < m_{\mathrm{jet}}< 234 \gev$  &  $ 162 < m_{\mathrm{jet}}< 265  \gev$  \\
\hline
\end{tabular}
\begin{tablenotes}
\item [a] \footnotesize{The optimal $z_\text{cut}$ found is near the ``standard'' value of 0.1, but much smaller values of $D_\text{cut}$ are found (the original value was 0.5).  This is due to a trade-off between the pruning and mass cut parameters.  With wide mass windows and high efficiencies, it turns out to be better to ``over-prune''.  The fact that $D_\text{cut}$ decreases from the 20\% efficiency point to the 50\% point is likely an artifact of the low resolution of the parameter scan (cf. Fig. \ref{fig:hadroniccomp:mistag_eff}).}
\item[b] {The variant of the ATLAS tagger in these proceedings is based on a cut on the likelihood value from TMVA and hence parameter values are not applicable.} 
\end{tablenotes}
\caption{Optimised parameters at different working points for different top-taggers.\label{tab:taggerpars}}
\end{threeparttable}
\end{center}
\end{table*}

The ATLAS tagger\footnote{The ATLAS studies of the variables used in this tagger only became public after BOOST2010~\cite{ATL-PHYS-PUB-2010-008}.} relies on $m_{\mathrm{jet}}$, $m_W$\footnote{The $W$ boson mass is defined as
 the lowest pairwise mass among the three subjets obtained by undoing the two 
last stages of the $k_T$ clustering.} and a variant of the first three 
splitting scales that gives dimensionless 
observables\footnote{$z_{cut} \equiv \frac{d_{cut}}{d_{cut}+m_{\mathrm{jet}}^2}$, where 
$d_{cut}$ is the $k_T$ distance between the merging subjets and $m_{\mathrm{jet}}$ is the 
mass of the merged jet.}. In order to ease subsequent analysis, we used a 
projective (one dimensional) likelihood estimator  to discriminate signal from
 background events. The likelihood classifier was built with the {\tt TMVA} 
toolkit~\cite{hocker2007tmva}. The Thaler/Wang tagger makes use of $m_{\mathrm{jet}}$, 
$m_W$ and a dimensionless energy sharing observable among the last two 
subjets\footnote{$z_{cell} \equiv \frac{min(E_1,E_2)}{E_1+E_2}$, where $E_i$ is
 the energy of the $i^{ith}$ subjet when undoing the last stage of the $k_T$ 
clustering process~\cite{Thaler:2008ju}.}.  In this study, we optimised rectangular cuts on the variables used by the  Thaler/Wang algorithm with {\tt TMVA} for the classification of events.  The resulting efficiencies as a function 
of jet $p_T$ are shown in Fig.~\ref{fig:hadroniccomp:tagrate_pt}.  
The efficiencies are relatively flat  for  $p_T \gtrsim 500 \gev$ after a turn-on for lower $p_T$. We also indicate the maximum possible efficiency after applying the $m_{\mathrm{jet}} >120\,\GeV$ cut in the same figure.

To optimise the  Hopkins tagger and its close cousin, the CMS tagger, we varied the lower cut for the jet mass, $m_{\text{jet}}$, and the cut window for $m_{\text{min}}$, $m_{W}$, yielding the curves in Fig.~\ref{fig:hadroniccomp:mistag_eff}. In addition, the  two taggers are compared in Fig. \ref{fig:hadroniccomp:tagrate_pt} for two working points chosen to yield 20\% and 50\% overall top-tagging efficiency. We find the tag rate to be relatively flat for $p_T \gtrsim 500 \GeV$ after a steep turn-on for lower 
$p_T$. The small $p_T$ dependence at higher $p_T$ in the taggers which do not employ grooming is further reduced in these grooming-based taggers.

We finally consider a top-tagger that employs pruning to groom the jets (described in detail in Section~\ref{sec:had:tools:groom}). 
For the purposes of this study, we included an additional step: To identify the $W$ boson subjet, the final jet is unclustered to three subjets (by undoing the
last merging) and the minimum-mass pairing is chosen to be the $W$ boson, as in the CMS tagger.

To generate the pruning tagger efficiency curves in Fig.~\ref{fig:hadroniccomp:mistag_eff},
the parameters $z_\text{cut}$ and $D_\text{cut}$ are scanned over the ranges 0.01--0.2 and (0.1--0.85)$\times (2 m/p_T)_\text{jet}$.  We then scan the cuts on the jet and $W$ boson subjet masses, with the only constraint being that the top jet mass is always required to be greater than 120 \gev.
We define two working points, that yield an average efficiency of 20\% and 50\%. The tagger parameters of both working points are given in Table~\ref{tab:taggerpars}. The tagging rates for signal and background as functions of anti-$k_T$ jet $p_T$ are shown in Fig.~\ref{fig:hadroniccomp:tagrate_pt}. The tag rates are relatively flat for $p_T \gtrsim 400 \gev$, after a turn-on for lower $p_T$. 

In general all grooming-based taggers that we tested have a flatter efficiency above $p_T$ of 400 GeV than the ungroomed approaches. This reflects the relative stability of the groomed variables as a function of $p_T$. Splitting scales, in particular, are sensitive to the $p_T$ of the initial jets, however groomed 
masses correspond closely to physical quantities and hence are Lorentz-boost invariant.

The overall mistag rates for the different taggers at the different working points are summarised in Table~\ref{tab:taggersummary}. Statistical errors are quoted for all measurements.  

Before we discuss these results in more detail, it is useful to discuss the reliability of such performance estimates. In Section~\ref{sec:had:robust} we found that the distribution of jet substructure observables is rather sensitive to the choice of parton shower and underlying event model tune. To quantify the effect on the tagging performance, we measured the performance of the taggers with the parameters of Table~\ref{tab:taggerpars} on a second sample generated with {\sc PYTHIA}. As might be expected from the results of Section~\ref{sec:had:robust} and earlier studies~\cite{Kaplan:2008ie}, we find that this ad hoc choice has a profound impact on the performance. The performance of all taggers as measured on the {\sc PYTHIA} sample is significantly better than for the default {\sc HERWIG} samples, with the rate of fake tags in the di-jet sample (for equal efficiency) dropping by up to a factor 2. If we assume the difference between both samples is an indication of the systematic error, the absolute value of the fake rate is uncertain to a level that makes comparison very hard, if not impossible. In Section~\ref{sec:had:robust}, we found, moreover, that even simple detector effects have a profound impact on substructure observables. The absence of a detailed detector simulation thus further undermines the reliability of the absolute performance measurements.

The relative performance of the taggers, however, is not affected by this large systematic error. The use of common benchmark samples ensures that the tagging approaches are compared on a level playing field. When we compare the {\sc PYTHIA} and {\sc HERWIG} results we indeed find that, despite the large changes in the absolute fake rate, the relative performance of the different taggers is conserved. To enable direct comparison with the existing taggers, we recommend that future taggers be tested using these samples.

We can then proceed to a discussion of the relative performance of the taggers. For the 20\% working point it is clear that the grooming based taggers perform strongly, suppressing the background by a factor of 20--100. For the samples we chose, the pruning approach performs best. The ungroomed tagging approaches are more competitive at the 50\% working point, which is often at the limit of the applicable range for the grooming-based approaches. It can be seen that the pruning-based approach actually performs worst at this working point. 

This seems to be the reflection of the fact that grooming approaches produce a narrow top mass peak, typically containing around 60\% of the signal for top jets. To produce an overall efficiency of around 50\%, in combination with the $m_{\mathrm{jet}} > 120\,\mathrm{GeV}$ requirement, we must then choose a large mass window. This partly negates the advantages of the grooming approaches and leads to worse relative performance compared to techniques without grooming.

\begin{table}[htb!]
\begin{center}
\begin{tabular}{|c|c|c|c|c|}
\hline
        \multicolumn{5}{|c|}{{\sc HERWIG} results}  \\ \hline
               &  eff.  & mistag & eff.  & mistag \\
Tagger         &  ( \% ) & rate ( \% ) & ( \% ) & rate ( \% ) \\ \hline
Hopkins      & 20  &  0.4 $\pm$ 0.02  & 50  &  4.9 $\pm$ 0.06 \\ 
CMS          & 20  &  0.4 $\pm$ 0.02  & 50  & 5.2 $\pm$ 0.06  \\ 
Pruning      & 20  &  0.3 $\pm$ 0.02  & 50  &  7.6 $\pm$ 0.08  \\
ATLAS        & 20  &  0.7 $\pm$ 0.02  & 50  &  4.6 $\pm$ 0.06  \\ 
T/W  & 20  &  1.5 $\pm$ 0.04  & 50 & 6.0 $\pm$ 0.07 \\ 
\hline
               \multicolumn{5}{|c|}{{\sc PYTHIA} results}  \\  \hline
              &  eff.  & mistag & eff.  & mistag \\
Tagger         &  ( \% ) & rate ( \% ) & ( \% ) & rate ( \% ) \\ \hline
Hopkins      &  20       &   0.2  $\pm$ 0.01    &  47     &   3.2  $\pm$ 0.05    \\ 
CMS          &  22   &  0.3 $\pm$ 0.01 & 49  &  3.5 $\pm$ 0.05 \\ 
Pruning      &  19   & 0.2  $\pm$ 0.01  &   49    & 4.5   $\pm$ 0.06     \\
ATLAS        &  18   &  0.5 $\pm$ 0.02 & 49 &  3.1 $\pm$ 0.05 \\ 
T/W  &  18   &  0.8 $\pm$ 0.02 & 57  &  7.0 $\pm$ 0.08 \\ 
\hline
\end{tabular}
\end{center}
\caption{Summary of tagging efficiency and mistag rates at different working points for a number of top-taggers. To facilitate comparison the parameters are chosen such that all taggers run at 20 or 50 \% efficiency for the default {\sc HERWIG} samples. The same parameters are used on the {\sc PYTHIA} Perugia0 sample. Statistical errors on the mis-tag rate are indicated. The efficiency numbers have uncertainties of 0.1 \% \label{tab:taggersummary}}
\end{table}



\section{Conclusions}
\label{sec:conclusions}

At the LHC, many of the particles that we have considered heavy so far ($W$ and $Z$ bosons, 
the top quark, the Higgs boson and possible BSM particles in the same mass range) will be 
produced with a transverse momentum that greatly exceeds their mass. The topologies that 
form in the decay of such highly boosted particles are expected to play an important 
role in searches for BSM physics. The BOOST2010 workshop brought together leading 
theorists and experimentalists in this field of study. In this paper, we present the 
report from the hadronic working group.

Many groups have studied the use of boosted objects in a range of Standard Model and new physics scenarios, demonstrating that these topologies can increase the experiments' potential in many different areas of the LHC physics programme, from searches for the Higgs boson to the reconstruction of SUSY cascade decays and heavy resonances. 
We hope that the review section may provide a starting point for people interested in this exciting subject.
Two further review sections should serve as an inventory of the literature on the subject and provide links to the relevant results obtained in previous experiments. 

For experiments to benefit fully from the opportunities offered by boosted objects an extensive set of novel tools is required. 
We have prepared a number of samples to study the particle-level performance of these tools. We propose these be used as a benchmark for future analyses.

Jet grooming methods like pruning, trimming and filtering are particularly promising and the full deployment of these tools in the LHC experiments should be pursued actively. 
The comparison of the mass distributions for {\em raw}, ungroomed jets and after applying the pruning, 
filtering or trimming procedure reveals a clear improvement in the mass resolution for composite 
fat-jets formed in the hadronic decay of boosted top quarks. Consequently, the 
signal-to-background ratio in a given mass window is greatly improved. The different approaches to 
jet grooming yield rather similar results for the observables studied.

We have furthermore investigated the sensitivity of jet substructure observables 
to the Monte Carlo description. This study demonstrates that variations in the 
parton shower model, the underlying event activity or the detector model can have 
a non-negligible impact, especially on the jet mass. The result underlines the 
need for standard benchmark samples. Jet grooming is a very effective means to 
reduce the dependence of the jet mass on soft unrelated activity (underlying 
event, pile-up). The splitting scales are found to exhibit a much less pronounced 
sensitivity to such activity. These first results call for a much more detailed 
evaluation of the systematics in a realistic experimental environment.

Finally, we have compared different top tagging algorithms in identical 
conditions, i.e using common samples, primary jet reconstruction algorithms 
and performance estimators. We find boosted hadronic top decays can be tagged with an efficiency of well over 50\%, while rejecting jets from QCD background by over a factor 20. For 20\% efficiency the groomed taggers outperform their ungroomed counterparts, reaching a QCD jet rejection of over a factor of 200.

\section*{Acknowledgement}

We thank UK's IPPP, IoP, STFC Particle Physics Department, Dalitz Institute for Theoretical Physics at Oxford Physics department and Oxford sub-department of Particle Physics for making Boost2010 possible.
We thank the IT teams at Oxford, UW and LPTHE for the facilities. 
We especially thank Chris Causer for his help with the Oxford SVN system.

Previous discussions of the jet substructure tools described here occurred
during the Joint Theoretical-Experimental Workshop on Jets and Jet
Substructure at the LHC held at the University of Washington in January,
2010 and supported in part by the DOE under Task TeV of contract
DE-FG02-96ER40956.

We acknowledge the support of the STFC and the Royal Society (in particular, International Joint Projects 2008/2), United Kingdom; DOE and NSF, United States
of America.

GPS would like to thank Matteo Cacciari, Sebastian Sapeta and Gregory Soyez 
for collaboration on the simulation framework used to generate the events 
produced here. GPS acknowledges funding from French Agence Nationale
de la Recherche, grant ANR-09-BLAN-0060.

MV is funded under the Ramon y Cajal programme of the Ministerio de Educaci\'{o}n y Ciencia, Spain.

\bibliographystyle{unsrt}
\bibliography{boost2010}{}

\end{document}